\documentclass[a4paper]{JHEP3}

\usepackage{amsmath}
\usepackage{amssymb}
\usepackage{amsfonts}
\usepackage{bbm}
\usepackage{graphics}
\usepackage{graphicx}
\usepackage{subfigure}
\usepackage{hhline}
\usepackage{longtable}
\usepackage{multirow}
\usepackage{dsfont}

\newcommand{\cE}{\mathcal{E}}

\newcommand{\cC}{\mathcal{C}}

\newcommand{\cN}{\mathcal{N}}

\newcommand{\cM}{\mathcal{M}}

\def\IZ{{\mathbb Z}}
\def\IR{{\mathbb R}}

\def\IC{{\mathbb C}}
\def\IP{{\mathbb P}}

\newcommand{\tr}{{\rm Tr}}
\newcommand{\sgn}{{\rm sgn}}

\newcommand{\p}{\partial}
\newcommand{\tb}{\overline{\tau}}
\newcommand{\ov}{\overline}
\newcommand{\wh}{\widehat}

\newcommand{\half}{\frac12}

\newcommand{\qu}{\underline{q}}

\newcommand{\be}{\begin{equation}}
\newcommand{\ee}{\end{equation}}
\newcommand{\ba}{\begin{aligned}}
\newcommand{\ea}{\end{aligned}}

\newcommand{\ben}{\begin{eqnarray}}
\newcommand{\een}{\end{eqnarray}}

\title{Wall-crossing holomorphic anomaly and mock modularity of multiple M5-branes}
\author{Murad Alim$^{1}$,  Babak Haghighat$^{2,3}$, Michael Hecht$^4$, Albrecht Klemm$^3$, Marco Rauch$^3$ and Thomas Wotschke$^3$
\\
\\ $^1$Jefferson Physical Laboratory, Harvard University, Cambridge, MA, 02138, USA
\\ $^2$Institute for Theoretical Physics and Spinoza Institute, Utrecht University, 3508 TD Utrecht, The Netherlands
\\ $^3$Bethe Center for Theoretical Physics, University of Bonn, Nussallee 12, D-53115 Bonn, Germany
\\ $^4$Arnold Sommerfeld Center for Theoretical Physics, LMU, Theresienstr. 37, D-80333 Munich, Germany
}

\abstract{Using wall-crossing formulae and the theory of mock modular forms we derive a holomorphic anomaly equation for the modified elliptic genus of two M5-branes wrapping a rigid divisor inside a Calabi-Yau manifold. The anomaly originates from restoring modularity of an indefinite theta-function capturing the wall-crossing of BPS invariants associated to D4-D2-D0 brane systems. We show the compatibility of this equation with anomaly equations previously observed in the context of $\mathcal{N}=4$ topological Yang-Mills theory on $\mathbbm{P}^2$ and E-strings obtained from wrapping M5-branes on a del Pezzo surface. The non-holomorphic part is related to the contribution originating from bound-states of singly wrapped M5-branes on the divisor. We show in examples that the information provided by the anomaly is enough to compute the BPS degeneracies for certain charges. We further speculate on a natural extension of the anomaly to higher D4-brane charge.}

\preprint{BONN-TH-2010-13\\LMU-ASC 102/10}

\begin{document}

\section{Introduction}

The study of background dependence of physical theories has been a rich source of insights. Understanding the change of correlators as the background parameters are 
varied supplemented by boundary data can be sufficient to solve the theory. A class of theories where the question of background dependence can be sharply stated 
are topological field theories. Correlators in topological theories typically have holomorphic expansions near special values of the background moduli. The expansion 
coefficients can be given precise mathematical meaning as topological invariants of the geometrical configuration contributing to the topological non-trivial 
sector of the path integral. Physically the expansion often captures information of the degeneracies of BPS states of theories related to the same geometry.

An example of this is the topological A-model \cite{Witten:1988xj}, with a Calabi-Yau three-fold (CY) $X$ as target space, which in a large volume limit counts holomorphic maps from the world-sheet into $H_2(X,\mathbbm{Z})$ and physically captures the degeneracies of BPS states coming from an M-theory compactification on $X$ \cite{Gopakumar:1998ii,Gopakumar:1998jq}. Another example is the modified elliptic genus of an M5-brane wrapping a complex surface $P$,\footnote{In the following we will use the terms surface, divisor and four-cycle (of a CY) interchangeably when the context is clear.} which was related in ref.~\cite{Minahan:1998vr} to the partition function of topologically twisted $\mathcal{N}=4$ Yang-Mills theory \cite{Vafa:1994tf}, which computes generating functions of Euler numbers of moduli spaces of instantons.  This same quantity was shown in ref.~\cite{Gaiotto:2006wm} to capture the geometric counting of degeneracies of systems of D4-D2-D0 black holes associated to the MSW string \cite{Maldacena:1997de}.

In both cases the topological theories enjoy  duality symmetries. $T$-duality acting on the K\"ahler moduli on $X$ in the topological string case and $S$-duality for the $\mathcal{N}=4$ SYM theory acting on the gauge coupling $\tau=\frac{4 \pi i}{g^2} + \frac{\theta}{2 \pi}$. The former
symmetry extends by mirror symmetry and both might extend to $U$-duality groups. Both symmetries can be conveniently expressed in the language of modular forms.

The holomorphic expansions of the topological string correlators are given in the moduli spaces of families of theories. Fixing a certain background corresponding to a certain point in the moduli space, the topological correlators are expected to be holomorphic expansions. In refs.~\cite{Bershadsky:1993ta,Bershadsky:1993cx} holomorphic anomaly equations governing topological string amplitudes were derived showing that this is not the case and hence the correlators suffer from background dependence.\footnote{The anomaly relates correlators at a given genus to lower genera thus providing a way to solve the theory. Using a polynomial algorithm \cite{Yamaguchi:2004bt,Grimm:2007tm,Alim:2007qj} and boundary conditions \cite{Huang:2006si} this can be used to compute higher genus topological string amplitudes on compact CY \cite{Huang:2006hq} manifolds and solve it on non-compact CY \cite{Haghighat:2008gw,Alim:2008kp}.} In ref.~\cite{Witten:1993ed} a background independent meaning was given to the correlators, stating that the anomaly merely reflects the choice of polarization if the partition function is considered as a wave function only depending on half of the variables of some phase space which has a natural geometric meaning in this context.

This anomaly is also manifest in a failure of target space duality invariance of the holomorphic expansion which can only be restored at the expense of holomorphicity as shown in ref.~\cite{Aganagic:2006wq}.\footnote{Following the anomaly reformulation of refs.~\cite{Dijkgraaf:2002ac,Verlinde:2004ck}, see also \cite{Gunaydin:2006bz}.}  A similar story showed up in $\mathcal{N}=4$ topological $U(2)$ SYM theory on $\mathbbm{P}^2$ \cite{Vafa:1994tf}, where it was shown that different sectors of the partition function need a non-holomorphic completion which was found earlier in ref.~\cite{Zagier:1975} in order to restore $S$-dualtiy invariance.  An anomaly equation describing this non-holomorphicity was expected \cite{Vafa:1994tf} in the cases where $b_{2}^+(P)=1$. In these cases holomorphic deformations of the canonical bundle are absent. The non-holomorphic contributions were associated with reducible connections $U(n) \rightarrow U(m)\times U(n-m)$ \cite{Vafa:1994tf,Minahan:1998vr}. In ref.~\cite{Minahan:1998vr} this anomaly was furthermore related to an anomaly appearing in the context of E-strings \cite{Minahan:1997ct}. These strings arise from an M5-brane wrapping a del Pezzo surface  $\mathcal{B}_9$, also called $\frac{1}{2}$K3. The anomaly in this context was related to the fact that $n$ of these strings can form bound-states of $m$ and $(n-m)$ strings. Furthermore, the anomaly could also be related to the one appearing in topological string theory.
	
The anomaly thus follows from the formation of bound-states. Although the holomorphic expansion would not know about the contribution from bound-states, the restoration of duality symmetry forces one to take these contributions into account. The non-holomorphicity can be understood physically as the result of a regularization procedure. The path integral produces objects like theta-functions associated to indefinite quadratic forms which need to be regularized to avoid divergences. This regularization breaks the modular symmetry, restoring the symmetry gives non-holomorphic objects. The general mathematical framework to describe these non-holomorphic completions is the theory of mock modular forms developed by Zwegers in ref.~\cite{Zwegers:2002}.\footnote{See \cite{Zagier:2007,MR2555930} and app.~\ref{sec:mock} for an introduction and overview.} A mock modular form $h(\tau)$ of weight $k$ is a holomorphic function which becomes modular after the addition of a function $g^*(\tau)$, at the cost 
of losing its holomorphicity. Here, $g^*(\tau)$ is constructed from a modular form $g(\tau)$ of weight $2-k$, which is referred to as shadow.

Another manifestation of the background dependence of the holomorphic expansions of the  topological theories are wall-crossing phenomena associated to the enumerative content of the expansions. Mathematically, it is known that Donaldson-Thomas invariants jump on surfaces with $b_2^+(P)=1$, see \cite{Gottsche:1998} and references therein, for related physical works see for example refs.~\cite{Moore:1997pc,Losev:1997tp}. On the physics side wall-crossing refers to the jumping of the degeneracies of BPS states when walls of marginal stability are crossed. These phenomena were observed in the jumps of the soliton spectrum of two-dimensional theories  \cite{Cecotti:1992rm} and were an essential ingredient of the work of Seiberg and Witten \cite{Seiberg:1994rs} in four-dimensional theories. Recent progress
was triggered by formulae relating the degeneracies on both sides of
the walls, which were given from a supergravity analysis in
refs.~\cite{Denef:2000nb,Denef:2007vg} and culminated in a mathematical rigorous formula of Kontsevich and Soibelman (KS) \cite{KS:2008}, which could also be derived 
from continuity of physical quantities in refs.~\cite{Gaiotto:2008cd,Cecotti:2009uf} (See also refs.~\cite{Gaiotto:2009hg,Gaiotto:2010be,Cecotti:2010fi}). The 
fact that the holomorphic anomaly describes how to transform the counting functions when varying the background moduli, which
in turn changes the degeneracy of BPS states, suggests that
non-holomorphicity and wall-crossing are closely related. In
fact the failure of holomorphicity can be traced back to the boundary
of the moduli space of the geometrical configuration, where the
latter splits in several configurations with the same
topological  charges. Mock modularity was used in a physical context studying the wall-crossing of degeneracies of $\mathcal{N}=4$ dyons\footnote{See for example ref.~\cite{Dabholkar:2007zz} and references therein for more details.} in ref.~\cite{Dabholkar:2010}. In the context of $\mathcal{N}=2$ supersymmetric theories the application of ideas related to mock modularity was initiated in ref.~\cite{Manschot:2009ia} and further pursued in refs.~\cite{Manschot:2010xp,Manschot:2010nc,Bringmann:2010sd}. These motivated parts of our work.\footnote{Further physical appearances of mock modularity can be found for example in refs.~\cite{Eguchi:2008gc,Eguchi:2009cq,Eguchi:2010ej,Cheng:2010pq,Troost:2010ud}.}

In this paper we study the relation between wall-crossing and non-holomorphicity and relate the appearance of the two. A central role is played by a wall-crossing formula by G\"ottsche \cite{Gottsche:1999}, where the K\"ahler moduli dependence of a generating function of Euler numbers of stable sheaves is given in terms of an indefinite theta-function due to G\"ottsche and Zagier \cite{Gottsche:1998}. We show that this formula is equivalent to wall-crossing formulae of D4-D2-D0 systems in type IIA. The latter can be related to the (modified) elliptic genus of multiple M5-branes wrapping a surface. Rigid surfaces are subject to G\"ottsche's wall-crossing formula. Using ideas of Zwegers \cite{Zwegers:2002}, we translate the latter into a holomorphic anomaly equation for two M5-branes wrapping the surface/divisor. We show that this anomaly equation is the equation which was found in the context of $\mathcal{N}=4$ SYM \cite{Vafa:1994tf} and E-strings \cite{Minahan:1997ct,Minahan:1998vr}. We further propose the generalization of the anomaly equation for higher wrappings and comment on its implications for the wall-crossing of multiple D4-branes.

The organization of this work is as follows. In section \ref{sectwo} we review the effective description of physical theories obtained from wrapping $n$ M5-branes on rigid divisors. Depending on the perspective, this is either described in terms of the MSW CFT with (0,4) world-sheet supersymmetry \cite{Maldacena:1997de} or by $U(n)$ $\mathcal{N}=4$ topological SYM \cite{Vafa:1994tf}. Both cases admit a decomposition into theta-functions, the latter carry a dependence on the chosen K\"ahler class which determines the split into right- and left-movers. We recall the equivalent type IIA D4-D2-D0 brane description of the BPS states of the M5-branes. We continue with outlining the $\mathcal{N}=4$ Vafa-Witten theory and recall how bound-states of several E-strings cause an anomaly of the corresponding partition function.

In section 3 we first show the equivalence of the Kontsevich-Soibelman wall-crossing formula and a formula found by G\"ottsche in terms of an indefinite theta-function which captures the wall-crossing of the generating function of Euler numbers of moduli spaces of stable sheaves on a complex surface $P$ with $b_2^+(P)=1$. Physically this corresponds to the D4-D2-D0 bound-state description. We remedy the non-modularity of the indefinite theta-function using the ideas of Zwegers \cite{Zwegers:2002}. With the results at hand we prove a holomorphic anomaly equation for two M5-branes wrapping the divisor.

In section 4 we apply the wall-crossing formula to compute the elliptic genus for several examples of surfaces with $b_2^+(P)=1$. Moreover, we provide the form of the holomorphic anomaly equation in the higher charge case that is compatible with the simple form appearing in the context of E-strings. We furthermore explore the possibility of the anomaly having its origin in a choice of contour while doing the Fourier expansion of a meromorphic Jacobi form paralleling the reasoning in $\mathcal{N}=4$ dyon wall-crossing \cite{Dabholkar:2010}.

Section 5 presents our conclusions, points out open problems and suggestions for future work. In the appendices we summarize several details.




\section{Effective descriptions of wrapped M5-branes}\label{sectwo}

In this section we review the effective descriptions of M5-branes wrapping a complex surface $P$ as well as previous appearances of the holomorphic anomaly which will be derived in the next section.  The world-volume theory of M5-branes can have either a two-dimensional CFT description in terms of the (MSW) CFT \cite{Maldacena:1997de} or a four-dimensional description giving the $\mathcal{N}=4$ topologically twisted Yang-Mills theory of Vafa and Witten \cite{Vafa:1994tf}. In the latter theory it was observed \cite{Vafa:1994tf} that a non-holomorphicity \cite{Zagier:1975} had to be introduced in order to restore $S$-dualtiy, the resulting holomorphic anomaly was related in ref.~\cite{Minahan:1998vr} to an anomaly~\cite{Minahan:1997ct} appearing in the context of E-strings. The anomaly was conjectured to take into account contributions coming from reducible connections in $\mathcal{N}=4$ SYM theory. In ref.~\cite{Minahan:1998vr} it was related to the curve counting anomaly \cite{Bershadsky:1993cx} and was given the physical interpretation of taking into account the bound-state contribution of E-strings. Later we will show that the contributions from bound-states as a cause for non-holomorphicity will persist more generally for the class of surfaces we will be studying. In our work we investigate the (generalized/modified) elliptic genus which captures the content of the CFT description of the M5-branes \cite{Minahan:1998vr,Maldacena:1999bp} and its relation to D4-D2-D0 systems \cite{Gaiotto:2006wm,deBoer:2006vg,Kraus:2006nb,Denef:2007vg,Gaiotto:2007cd,Manschot:2007ha} and the associated counting of black holes which has been intensively studied (e.g.~in ref.~\cite{Dabholkar:2005dt}).  Our goal is to show that wall-crossing in D4-D2-D0 systems leads to an anomaly equation which coincides with the anomalies found before and hence our work complements in some sense this circle of ideas.

\subsection{The elliptic genus and D4-D2-D0 branes}
\label{sec:mswcft}
In the following we will start with the $2d$ CFT perspective of the M5-brane world-volume theory.
We want to study BPS states that arise in the context of an M-theory compactification on a Calabi-Yau manifold $X$ with $r$ M5-branes wrapping a complex surface (or a four-cycle) $P$,  and extended in $\mathbbm{R}^{1,3} \times S^1$. Considering $P$ to be small compared to the M-theory circle, the reduction of the world-volume theory of the M5-brane is described by a $(1+1)$-dimensional $(0,4)$ MSW CFT \cite{Maldacena:1997de}.\footnote{The target space sigma model description of which was given in ref.~\cite{Minasian:1999qn}, for more details see ref.~\cite{Guica:2007wd} and references therein. In the following we will be concerned with the natural extension of the analysis of the degrees of freedom to $r$ M5-branes.} The BPS states associated to the string that remains after wrapping the M5-branes on $P$ are captured by a further compactification on a circle. They are counted by the partition function of the world-volume theory of the M5-branes on $P\times T^2$ \cite{Minahan:1998vr}. The effective CFT description will thus exhibit invariance under the full SL$(2,\IZ)$ symmetry of the $T^2$. Furthermore, excitations of the M5-branes will induce M2-brane charges corresponding to the flux of the self-dual field strength of the M5-brane world-volume theory. In addition, the momentum of the M2-branes along the M-theory circle will give rise to a further quantum number. As a result BPS states of the effective two-dimensional description will be labeled by the class of the divisor the M5-branes wrap, the M2-brane charges and by the momentum along $S^1$. In a type IIA setup, $r$ times the class of the divisor will correspond to D4-brane charge, the induced M2-brane charge corresponds to D2-brane charge and the momentum to D0-brane charge. Choosing a basis $\Sigma_A\, , A=1,\dots, b_4(X)$ of $H_4(X,\IZ)$, the charge vector will be given by\footnote{See appendix \ref{Dbound} for details.}
\be
\Gamma = (Q_6,Q_4,Q_2,Q_0) = r(0,p^A, q_A, q_0),
\ee
where the $Q_p$ are the D$p$-brane charges and $r$ is the number of coincident M5-branes wrapping the divisor specified by $p^A$.  A priori the set of all possible induced D2-brane charges, or equivalently of $U(1)$ fluxes of the world-volume of the M5-brane would be in one-to-one correspondence with $\Lambda_P = H^2(P,\mathbbm{Z})$ which is generically a larger lattice than $\Lambda=i^*H^2(X,\IZ)$, where $ i: P \hookrightarrow X$, however the physical BPS states are always labeled by the smaller lattice $\Lambda$. The metric $d_{AB}$ on $\Lambda$ is given by
\be
d_{AB} = - \int_P \alpha_A \wedge \alpha_B,
\ee
where $\alpha_A$ is a basis of two-forms in $\Lambda$, which is the dual basis to $\Sigma_A$ of $H_4(X,\IZ)$.  In order to obtain a generating series of the degeneracies of those BPS states one has to sum over directions along $\Lambda^{\perp}$ which is the orthogonal complement to $\Lambda$ in $\Lambda_P$ w.r.t.~$d_{AB}$ \cite{Gaiotto:2006wm}. \footnote{In general, the lattice $\Lambda \oplus \Lambda^{\perp}$ is only a sublattice of $H^{2}(P,\mathbbm{Z})$, because $\det d_{AB}\neq1$ in general, see for example ref.~\cite{Minasian:1999qn} and ref.~\cite{Denef:2007vg} for a more recent exposition. However, we will only be concerned with divisors $P$ with $b_2^+(P)=1$, such that $\det d_{AB}=1$.}

The partition function of the MSW CFT counting the BPS states is given by the modified elliptic genus\footnote{We follow the mathematics convention of not writing out explicitly the dependence on $\tb$ which will be clear in the context. Moreover, we denote $q= e^{2\pi i \tau}$ and $\tau=\tau_1+i\tau_2$. To avoid confusion without introducing new notation we will denote the charge vector of D2-brane charges by $\underline{q}$, its components by $q_A$.} \cite{Minahan:1998vr,Maldacena:1999bp}
\begin{equation} \label{zprime}
{Z'}_{\!\!\!P}^{(r)}(\tau,z)= \textrm{Tr}_{\mathcal{H}_{\rm{RR}}} \, (-1)^{F_{{\rm R}}} \, F_{{\rm R}}^2\, q^{L'_0-\frac{c_{\rm L}}{24}}\, \bar q^{\bar{L}'_0 -\frac{c_{\rm R}}{24}}e^{2\pi i z\cdot Q_{2}},
\end{equation}
where the trace is taken over the RR Hilbert space. Furthermore, vectors are contracted w.r.t.~the metric $d_{AB}$, i.e. $x \cdot y = x^A y_A = d_{AB}x^A y^B$. For a single M5-brane it was shown in ref.~\cite{deBoer:2006vg} that ${Z'}_{\!\!\!P}^{(1)}(\tau,z)$ transforms like a SL$(2,\IZ)$ Jacobi form of bi-weight $(0,2)$ due to the insertion of $F_{{\rm R}}^2$, we demand that the same is true for all $r$.

Following ref.~\cite{deBoer:2006vg} the center of mass momentum $\vec{p}_{\rm cm}$ for the system of $r$ M5-branes can be integrated out. In this way $L_0'$ and $\bar{L}_0'$ can be written in the form
\begin{equation}
  L_0' = \frac{1}{2} \vec{p}^{\, 2}_{\rm cm} + L_0, \quad \bar{L}_0' = \frac{1}{2} \vec{p}^{\, 2}_{\rm cm} + \bar{L}_0.
\end{equation}
This allows one to split up the center of mass contribution and rewrite formula (\ref{zprime}) as
\begin{eqnarray}
  {Z'}_{\!\!\!P}^{(r)}(\tau,z)
  & =    & \int d^3 p_{\rm cm} (q\bar{q})^{\frac{1}{2}\vec{p}_{\rm cm}^{\, 2}} Z^{(r)}_{P}(\tau,z) \nonumber \\
  & \sim & (\tau_2)^{-\frac{3}{2}}\, Z^{(r)}_{P}(\tau,z),
\end{eqnarray}
where $Z^{(r)}_{P}(\tau,z)$ is now a Jacobi form of weight $(-\frac{3}{2},\frac{1}{2})$ which we simply call elliptic genus for short in the following.

\subsubsection*{\it The decomposition of the elliptic genus}
The elliptic genus $Z_P^{(r)}(\tau, z)$ and equivalently the generating function of D4-D2-D0 BPS degeneracies is subject to a theta-function decomposition, which has been studied in many places, see for example refs.~\cite{Dabholkar:2005dt,Gaiotto:2006wm,deBoer:2006vg,Kraus:2006nb,Denef:2007vg}. This is ensured by two features of the superconformal algebra of the (0,4) CFT. One of these is that the $\tb$ contribution entirely comes from BPS states $| \qu \rangle$ satisfying
\begin{equation} \label{heatop}
    \left(\ov{L}_0-\frac{c_{\rm R}}{24}-\frac{r}{2}q_{{\rm R}}^2 \right) |\qu\rangle =0,
\end{equation}
the other one is the spectral flow isomorphism of the ${\cal N}=(0,4)$ superconformal algebra, which we want to recall for $r$ M5-branes here, building on refs.~\cite{deBoer:2006vg,Weist:2009}, see also \cite{Dabholkar:2005dt}. Proposition 2.9 of ref.~\cite{Weist:2009} describes the spectral flow symmetry by an isomorphism between moduli spaces of vector bundles on complex surfaces. The complex surface here is the divisor $P$ and the vector bundle configuration describes the bound-states of D4-D2-D0 branes. Within this setup the result of \cite{Weist:2009} translates for arbitrary $r$ to a symmetry under the transformations
\begin{eqnarray} \label{spflow}
    q_0 & \mapsto & q_0 - k \cdot \qu - \frac{1}{2} k \cdot k, \nonumber \\
    \qu   & \mapsto & \qu + k,
\end{eqnarray}
where $k \in \Lambda$. Physically these transformations correspond to monodromies around the large radius point in the moduli-space of the Calabi-Yau manifold \cite{Dabholkar:2005dt}. Denote by $\Lambda^*$ the dual lattice of $\Lambda$ with respect to the metric $r d_{AB}$. Keeping only the holomorphic degrees of freedom one can write
\begin{eqnarray}
    Z^{(r)}_{P}(\tau,z) & = & \sum_{Q_0; Q_A}\, d(Q,Q_0)\,e^{-2\pi i \tau Q_0}\, e^{2\pi i z\cdot Q_{2}} \nonumber \\
    ~                     & = & \sum_{q_0; \qu \in \Lambda^* + \frac{[P]}{2}} \,d(r,\qu,-q_0)\,e^{-2\pi i \tau r q_0}\, e^{2\pi i r z\cdot \qu},
\end{eqnarray}
where $d(r,\qu,-q_0)$ are the BPS degeneracies and the shift\footnote{In components, $[P]$ is given by $d_{AB}p^A$.} $\frac{[P]}{2}$ originates from an anomaly \cite{Freed:1999vc,Minasian:1997mm}. Now, spectral flow symmetry predicts \cite{deBoer:2006vg}
\begin{equation} \label{spflow2}
    d(r,\qu,-q_0) = (-1)^{r p \cdot k} d(r,\qu+k,-q_0+k\cdot \qu + \frac{k^2}{2}).
\end{equation}
Making use of this symmetry and the following definition
\begin{equation}
 \qu = k + \mu + \frac{[P]}{2}, \qquad \mu \in \Lambda^*/\Lambda, \qquad k \in \Lambda,
\end{equation}
one is led to the conclusion that the elliptic genus can be decomposed in the form
\begin{eqnarray}
    Z^{(r)}_{P}(\tau,z) & = & \sum_{\mu \,\in\, \Lambda^*/\Lambda} f^{(r)}_{\mu,J}(\tau) \theta^{(r)}_{\mu,J}(\tau,z), \label{thetadecomp}\\
    f^{(r)}_{\mu,J}(\tau) & = & \sum_{r \hat{q}_0\,\geq\,-\frac{c_{\rm L}}{24}} d^{(r)}_{\mu}(\hat{q}_0) e^{2\pi i\tau r \hat{q}_0}, \label{fdef}\\
    \theta^{(r)}_{\mu,J}(\tau,z)
   & =& \sum_{k\,\in\,\Lambda+\frac{[P]}{2}} (-1)^{r p \cdot (k+\mu)}e^{2\pi i \bar{\tau} r \frac{(k+\mu)_+^2}{2}}
      e^{2\pi i \tau r \frac{(k+\mu)_-^2}{2}} e^{2\pi i r z \cdot (k+\mu)}\label{SNTheta},
\end{eqnarray}
where $J\in\cC(P)$ and $\cC(P)$ denotes the K\"ahler cone of $P$ restricted to $\Lambda\otimes\IR$ and $\hat{q}_0 = -q_0 - \frac{1}{2} \qu^2$ is invariant under the spectral flow symmetry. The subscript $+$ refers to projection onto the sublattice generated by the K\"ahler form $J$ and $-$ is the projection to its orthogonal complement, i.e.
\be
k_+^2=\frac{(k\cdot J)^2}{J\cdot J},\quad k_-^2=k^2-k_+^2.
\ee

There are two issues here for the case of rigid divisors with $b_2^+(P) = 1$ on which we want to comment as this class of divisors is the
focus of our work. First of all note, that $q_0$ contains a contribution of the form\footnote{See appendix \ref{Dbound} for details.} $\frac{1}{2} \int_P F \wedge F$ where $F \in \Lambda_P$. Now, $F$ can be decomposed into $F = \qu + \qu_{\perp}$ with $\qu_{\perp} \in \Lambda^{\perp}$, which allows us to write \be \hat{q}_0 = \tilde{q}_0 + \frac{1}{2}\qu_{\perp}^2.\ee For $b_2^+(P) =1$ and $r=1$, the degeneracies $d(r,\mu,\tilde{q}_0)$ are independent of the choice of $\qu_{\perp}$ and moreover it was shown by G\"ottsche  \cite{Gottsche:1990} that
\be
\sum_{\tilde{q}_0} d(1,\mu,\tilde{q}_0) \, e^{2\pi i \tau \tilde{q}_0} = \frac{1}{\eta^{\chi(P)}}.
\ee
Then, for $r=1$ (\ref{fdef}) becomes
\be \label{frigid}
    f^{(1)}_{\mu,J}(\tau)
=\frac{\vartheta_{\Lambda^{\perp}}(\tau)}{\eta^{\chi(P)}(\tau)}, \qquad
\vartheta_{\Lambda^{\perp}}(\tau)  =  \sum_{\qu_{\perp} \in \Lambda^{\perp}} e^{i\pi\tau \qu_{\perp}^2}.
\ee

The second subtlety is concerned with the dependence on a K\"ahler class $J$. Due to wall-crossing phenomena we will find that $f^{(r)}_{\mu,J}(\tau)$ also depends on $J$. We expect that it has the following expansion ($\tilde{q}_0=\frac{d}{r}-\frac{c_{\rm L}}{24}$)
\begin{equation} \label{fconj}
    f^{(r)}_{\mu,J}(\tau) = (-1)^{r p \cdot \mu} \,\sum_{d\,\geq\,0}\bar{\Omega}(\Gamma; J) \, q^{d - \frac{r\chi(P)}{24}}.
\end{equation}
Here, the factor $(-1)^{r p \cdot \mu}$ is inserted to cancel its counterpart in the definition of $\theta^{(r)}_{\mu,J}$, which was only included to make the theta-functions transform well under modular transformations. The invariants $\bar{\Omega}(\Gamma;J)$ are rational invariants first introduced by Joyce \cite{Joyce:2006pf,Joyce:2008} and are defined as follows
\begin{equation} \label{bomega}
    \bar{\Omega}(\Gamma; J) = \sum_{m|\Gamma} \frac{\Omega(\Gamma/m; J)}{m^2},
\end{equation}
where $\Omega(\Gamma,J)$ is an integer-valued index of BPS degeneracies, given by \cite{Dabholkar:2005by}
\be
\Omega(\Gamma, J) = \frac{1}{2} \tr (2J_3)^2 (-1)^{2J_3},
\ee
where $J_3$ is a generator of the rotation group ${\rm Spin}(3)$.  Note, that for a single M5-brane $\bar{\Omega}$ and $\Omega$ become identical and independent of $J$.

\subsection{$\mathcal{N}=4$ SYM, E-strings and bound-states}\label{n4ym}
In the following we recall the relation \cite{Minahan:1998vr} of the elliptic genus of M5-branes to the $\mathcal{N}=4$ topological SYM theory of Vafa and Witten \cite{Vafa:1994tf}. Our goal is to relate the holomorphic anomaly equation which we will derive from wall-crossing in the next section to the anomalies appearing in the ${\cal N}=4$ context. We review moreover the connection of the anomaly to the formation of bound-states given in ref.~\cite{Minahan:1998vr}.

The $\mathcal{N}=4$ topological SYM arises by taking a different perspective on the world-volume theory of $n$ M5-branes on $P\times T^2$ considering the theory living on $P$ which is the $\mathcal{N}=4$ topologically twisted SYM theory described in ref.~\cite{Vafa:1994tf}. The gauge coupling of this theory is given by
\be
\tau=\frac{4 \pi i }{g^2}+ \frac{\theta}{2 \pi},
\ee
and is geometrically realized by the complex structure modulus of the $T^2$.  The partition function of this theory counts instanton  configurations by computing the generating functions of the Euler numbers of moduli spaces of gauge instantons \cite{Vafa:1994tf}.  $S$-dualtiy translates to the modular transformation properties of the partition function. The analogues of D4-D2-D0 charges are the rank of the gauge group, different flux sectors and the instanton number. 

In ref.~\cite{Minahan:1998vr} the relation is made between this theory and the geometrical counting of BPS states of exceptional strings obtained by wrapping M5-branes around a del Pezzo surface $\mathcal{B}_9$, also called $\frac{1}{2}$K3. This string is dual to the heterotic string with an $E_8$ instanton of zero size \cite{Ganor:1996mu,Seiberg:1996vs} and is therefore called E-string. In F-theory this corresponds to a
$\mathbbm{P}^1$ shrinking to zero size
\cite{Morrison:1996na,Morrison:1996pp,Witten:1996qb}. 
The geometrical study of the BPS states of this non-critical string was initiated in ref.~\cite{Klemm:1996hh} and further pursued in refs.~\cite{Lerche:1996ni,Minahan:1997ch,Minahan:1997ct}. In
ref.~\cite{Minahan:1998vr} the counting of BPS states of the 
exceptional string with increasing winding $n$ was related to 
the $U(n)$, $\mathcal{N}=4$ SYM partition functions. 

In the following we will use the geometry of ref.~\cite{Klemm:1996hh} which is an elliptic fibration over the Hirzebruch surface $\mathbbm{F}_1$, which in turn is a $\mathbbm{P}^1$ fibration over $\mathbbm{P}^1$.\footnote{The toric data of this geometry is summarized in appendix \ref{toricdata}.} We will denote by $t_E, t_F$ and $t_D$ the K\"ahler parameters of the elliptic fiber, the fiber and the base of $\mathbbm{F}_1$, respectively and enumerate these by $1,2,3$ in this order. We further introduce $\tilde q_a=e^{2\pi i \tilde{t}_a}\, , \,a=1,2,3$ the exponentiated K\"ahler parameters appearing in the instanton expansion of the A-model at large radius, which are also the counting parameters of the BPS states. 

Within this geometry we will be interested in the elliptic genus of M5-branes wrapping two different surfaces, one is a K3 corresponding to wrapping the elliptic fiber and the fiber of $\mathbbm{F}_1$, the resulting string is the heterotic string. The other possibility is to wrap the base of $\mathbbm{F}_1$ and the elliptic fiber corresponding to $\frac{1}{2}$K3 and leading to the E-string studied in refs.~\cite{Klemm:1996hh,Lerche:1996ni,Minahan:1997ch,Minahan:1997ct,Minahan:1998vr}. The two possibilities are realized by taking the limits $t_D,t_F\rightarrow i \infty$, respectively. The resulting surface in both cases is still elliptically fibered which allows one to identify the D4-D0 charges $n$ and $p$ with counting curves wrapping $n$-times the base and $p$-times the fiber of the elliptic fibration \cite{Minahan:1998vr}. The multiple wrapping is hence encoded in the expansion of the prepotential $F_0(\tilde q_1,\tilde q_2,\tilde q_3)$ of the geometry. In order to get a parameterization inside the K\"ahler cone of the K3 in which the corresponding curves in $H_2({\rm K3},\IZ)$ intersect with the standard metric of the hyperbolic lattice $\Gamma^{1,1}$, we define $t_1=\tilde{t}_1\, ,t_2=\tilde{t}_2-\tilde{t}_1$ and $t_3=\tilde{t}_3$ as well as the corresponding $q_1 = \tilde q_1$, $q_2 = \tilde q_2/\tilde q_1$ and $q_3=\tilde q_3$. Taking $q_2$ or $q_3 \rightarrow 0$, the multiple wrapping of the base is expressed by
\begin{equation}
F_0(t_1,t_a)=\sum_{n\geq 1}Z^{(n)}(t_1) q_a^n\, , \quad a=2\, \textrm{ or } \,3. 
\end{equation}
The $Z^{(n)}$ can be identified with the elliptic genus of $n$ M5-branes wrapping the corresponding surface after taking a small elliptic fiber limit \cite{Minahan:1998vr}. In this limit the contribution coming from the theta-functions (\ref{SNTheta}) reduce to $\tau_2^{-3/2} \left(\tau_2^{-1/2}\right)$ for the K3($\frac{1}{2}$K3) cases, these are the contributions of 3(1) copies of the lattice $\Gamma^{1,1}$ appearing in the decomposition of the lattices of K3($\frac{1}{2}$K3). Omitting these factors gives the $Z^{(n)}$ of weight $(-2,0)$ in both cases. The elliptic genera of wrapping $n$ M5-branes corresponding to $n$ strings are in both cases related recursively to the lower
wrapping. The nature of the recursion depends crucially on the ability of the strings to form bound-states.

\subsubsection*{\it The heterotic string, no bound-states} \label{hetstring}
The heterotic string is obtained from wrapping an M5-brane on the K3 by taking the $q_3\rightarrow 0$ limit. The heterotic string does not form bound-states and the recursion giving the higher wrappings in this case is the Hecke transformation\footnote{For a review on Hecke transformations see Zagier's article in \cite{1-2-3-modular}.} of $Z^{(1)}$ as proposed in ref.~\cite{Minahan:1998vr}. The formula for the Hecke transformation in this case is given by
\begin{equation}\label{eq:Hecke}
Z^{(n)} (t)= n^{w_{\rm L}-1} \sum_{a,b,d} d^{-w_L} Z^{(1)}\left(  \frac{a t+b}{d}\right) \, ,
\end{equation}  
with $ad=n$ and $b < d$ and $a,b,d \ge 0$. Which specializes for  $w_{\rm L}=-2$ and $n=p$, where $p$ is prime to
\begin{equation} 
Z^{(p)} (t)= \frac{1}{p^3} Z^{(1)} (p t) + \frac{1}{p} \left[Z^{(1)}\left(\frac{t}{p}\right) + Z^{(1)}\left(\frac{t}{p} +\frac{1}{p}\right) + \dots + Z^{(1)}\left(\frac{t}{p} +\frac{p-1}{p}\right) \right] \, .
\end{equation}
For example the partition functions for $n=1,2$ obtained from the instanton part of the prepotential of the geometry read
\begin{equation}
Z^{(1)}=-\frac{2 E_4 E_6 }{\eta^{24}} ,\quad Z^{(2)}=-\frac{E_4 E_6 \left(17 E_4^3+7 E_6^2\right)}{96 \eta^{48}}\, ,
\end{equation}
and are related by the Hecke transformation. Further examples of higher wrapping are given in the appendix \ref{app:K3}. The fact that the partition functions of higher wrappings of the M5-brane on the K3, which correspond to multiple heterotic strings, are given by the Hecke transformation was interpreted \cite{Minahan:1998vr} by the absence of bound-states. Geometrically, multiple M5-branes on a K3 can be holomorphically deformed off one another. This argument fails for surfaces with $b_2^+=1$ and in particular for $\frac{1}{2}$K3.

One reason that the higher $Z^{(n)}$ can be determined in such a simple way from $Z^{(1)}$ can be understood in topological 
string theory from the fact that the BPS numbers on K3 depend only on the intersection of a curve ${\cal C}^2=2 g-2$~\cite{Yau:1995mv}, and not on their class in $H_2(\text{K3},\mathbb{Z})$. 
This allows to prove (\ref{eq:Hecke}) to all orders in the limit of the topological string partition function under consideration 
by slightly modifying the proof in~\cite{MKPS}. Using the Picard-Fuchs system of the elliptic fibration one shows in the limit 
$q_3\rightarrow 0$ the first equality in the identity 
\begin{equation} 
\begin{split}
\frac{1}{2}\left(\frac{\partial}{\partial{t_2}}\right)^3 F_0|_{q_3\rightarrow 0}
&=\frac{ E_4(t_1) E_6(t_1) E_4(t_2)}{\eta(t_1)^{24} (j(t_1)-j(t_2))}\\
&= \frac{q_1}{q_1-q_2}+ E_4(t_2)-\sum_{d,l,k>0} l^3 c(k l) 
q_1^{kl} q_2^{ld}\ ,
\end{split}
\end{equation}
where $j=E_4^3/\eta^{24}$ and $c(n)$ are defined as \be-\frac{1}{2}Z^{(1)}=\sum_n c(n) q^n.\ee This equations shows 
two things. The BPS numbers inside the K\"ahler cone of K3 depend only on ${\cal C}^2=k l$ and all  
$Z^{(n)}$ are given by one modular form. The second fact can be  used as in \cite{MKPS} to establish that 
\be\frac{1}{2}\left(\frac{\partial}{\partial{t_2}}\right)^3 F_0|_{q_3\rightarrow 0}=
\sum_{n=0}^\infty F_n(t_1) q_2^n ,\ee where $F_n$ is the Hecke transform of $F_1$, i.e.~$n^3 F_n=F_1| T_n$. Using Bol's identity and restoring the $n^3$ factors yields (\ref{eq:Hecke}).

\subsubsection*{\it E-strings and bound-states}
The recursion relating the higher windings of the E-strings to lower winding,  developed in \cite{Minahan:1997ch,Minahan:1997ct,Minahan:1998vr} in contrast reads
\begin{equation}
\frac{\partial Z^{(n)}}{\partial E_2}=\frac{1}{24} \sum_{s=1}^{n-1} s(n-s) Z^{(s)}\,Z^{(n-s)}\, ,
\end{equation}
which becomes an anomaly equation, when $E_2$ is completed into a modular object  $\wh{E}_2$ by introducing a non-holomorphic part (see appendix \ref{sec:mock}). The anomaly reads:
\begin{equation}
\partial_{\bar{t}_1} \wh{Z}^{(n)}=\frac{i (\textrm{Im}\, t_1)^{-2}}{16 \pi} \, \sum_{s=1}^{n-1} s(n-s) \wh{Z}^{(s)} \wh{Z}^{(n-s)}\, ,
\end{equation}
and was given the interpretation \cite{Minahan:1998vr} of taking into account the contributions from bound-states. 
Starting from~\cite{Klemm:1996hh}
\begin{equation} 
Z^{(1)} =\frac{E_4 \sqrt{q}}{\eta^{12}},
\end{equation}
and using the vanishing of BPS states of certain charges one obtains recursively all $Z^{(n)}$ \cite{Minahan:1997ch,Minahan:1997ct,Minahan:1998vr}. E.g. the $n=2$ the 
contribution reads:
\begin{equation}
  \wh{Z}^{(2)}=\frac{q E_4 E_6}{12 \eta^{24}} + \frac{q \wh{E_2} E_4^2}{24 \eta^{24}}\, ,
\end{equation}
where the second summand has the form $\wh{E}_2 \left(Z^{(1)}\right)^2$ and takes into account the contribution from bound-states of singly wrapped M5-branes. 

A relation to the anomaly equations appearing in topological string theory \cite{Bershadsky:1993cx} was pointed out in ref.~\cite{Minahan:1998vr} and proposed for arbitrary genus in refs.~\cite{Hosono:1999qc,Hosono:2002xj}. The higher genus generalization reads \cite{Hosono:1999qc,Hosono:2002xj}:
\begin{equation}\label{eq:highergenus}
\frac{\partial Z^{(n)}_{g}}{\partial E_2}=\frac{1}{24} \sum_{g_1+g_2=g}\sum_{s=1}^{n-1} s(n-s) Z^{(s)}_{g_1}\, Z^{(n-s)}_{g_2}\,+ \frac{n(n+1)}{24} Z^{(n)}_{g-1}\, \, ,
\end{equation}
where the instanton part of the A-model free energies at genus $g$ is denoted by
$F_g(q_1,q_2,q_3)$, and $F_g(q_1,q_2\rightarrow 0,q_3)= \sum_{n\ge1} Z^{(n)}_g q_3^n$.
The $Z^{(n)}_{g}$ have the form \cite{Hosono:2002xj}
\begin{equation}
Z^{(n)}_{g}=P^{(n)}_{g}(E_2,E_4,E_6) \frac{q_1^{ n/2}}{\eta^{12 n}} \, ,
\end{equation}
where $P^{(n)}_{g}$ denotes a quasi-modular form of weight $2g+6n-2$.

\subsection{Generating functions from wall-crossing}
In the last section we have argued that the partition function of $\cN=4$ $U(r)$ Super-Yang-Mills theory suffers from a holomorphic anomaly
for divisors with $b_2^+(P) = 1$. In fact there exists another way to see the anomaly which
is also intimately related to the computation of BPS degeneracies encoded in the elliptic genus
and will be the subject of this section.
This method relies on wall-crossing formulas and originally goes back to G\"ottsche and Zagier \cite{Gottsche:1998,Gottsche:1999}.
In the physics context it has also been employed in \cite{Manschot:2010xp,Manschot:2010nc}.
It will be used in section \ref{sec:wcmm} to derive the elliptic genus for BPS states and their anomaly rigorously.
In the following presentation we will be very sketchy as we merely want to stress the main ideas.
We refer to section \ref{sec:wcmm} for details.

The starting point is the Kontsevich-Soibelman formula \cite{KS:2008} which describes the wall-crossing
of bound-states of D-branes. Specifying to the case of two M5-branes and taking the equivalent
D4-D2-D0 point of view the Kontsevich-Soibelman formula reduces to the primitive wall-crossing formula
\begin{equation} \label{pwc1}
    \Delta \Omega(\Gamma;J\rightarrow J')  = \Omega(\Gamma; J') - \Omega(\Gamma;J)= (-1)^{\langle \Gamma_1,\Gamma_2\rangle -1} \langle \Gamma_1,\Gamma_2 \rangle\, \Omega(\Gamma_1) \,\Omega(\Gamma_2),
\end{equation}
which describes the change of BPS degeneracies of a bound-state with charge vector $\Gamma = \Gamma_1 + \Gamma_2$,
once a wall of marginal stability specified by $J_W$ is crossed. The symplectic charge product $\langle \cdot, \cdot \rangle$ is defined by
\be
\langle\Gamma_1,\Gamma_2\rangle=-Q_6^{(1)}Q_0^{(2)}+Q_4^{(1)}\cdot Q_2^{(2)}-Q_2^{(1)}\cdot Q_4^{(2)}+Q_0^{(1)}Q_6^{(2)}.
\ee
Hence, for D4-D2-D0 brane configurations $\langle\Gamma_1,\Gamma_2\rangle$ is independent of the D0-brane charge.
Further, in eq.~(\ref{pwc1}) $\Gamma_1$ and $\Gamma_2$ are primitive charge vectors such that $\Omega(\Gamma_i)$ do not depend on the moduli.
Thus, the $\Gamma_i$ can be thought of as charge vectors with $r=1$ whereas $\Gamma$ corresponds to a charge vector with $r=2$.
Assuming, that the wall of marginal stability does not depend on the D0-brane charge, formula (\ref{pwc1}) can be translated into a generating series $\Delta f^{(2)}_{\mu,J\rightarrow J'}$ defined by
\be
    \Delta f_{\mu,J \rightarrow J'}^{(2)} = \sum_{d\geq 0} \Delta \bar{\Omega}(\Gamma;J\rightarrow J') \, q^{d-\frac{\chi(P)}{12}}.
\ee
Assuming that there exists a reference chamber $J'$ such that $\bar\Omega(\Gamma;J)=0$, this gives us directly an expression for $f^{(2)}_{\mu,J}$.

As it will turn out in the next section, $\Delta f^{(2)}_{\mu,J\rightarrow J'}$ is given in terms of an indefinite theta-function $\Theta^{J,J'}_{\Lambda,\mu}$, which contains the information about the decays due to wall-crossing as one moves from $J$ to $J'$. Indefinite theta-functions were analyzed by Zwegers in his thesis \cite{Zwegers:2002}. One of their major properties is that they are not modular as one only sums over a bounded domain of the lattice $\Lambda$ specified by $J$ and $J'$. However, Zwegers showed that by adding a non-holomorphic completion the indefinite theta-functions have modular transformation behavior and fall into the class of mock modular forms.\footnote{We review some notions in appendix \ref{sec:mock}.} Every mock modular form $h$ of weight $k$ has a shadow $g$, which is a modular form of weight $2-k$, such that the function
\begin{equation} \label{mock}
    \hat{h}(\tau) = h(\tau) + g^*(\tau)
\end{equation}
transforms as a modular form of weight $k$ but is not holomorphic. Here, $g^*$ is a certain transformation of the function $g$ that introduces a non-holomorphic dependence. Taking the derivative of $\hat{h}$ with respect to $\bar\tau$ yields a holomorphic anomaly given by the shadow
\begin{equation}
    \frac{\p \hat{h}}{\p \bar{\tau}} = \frac{\p g^*}{\p \bar{\tau}} = \tau_2^{-k} \overline{g(\tau)},
\end{equation}
where $\tau_2=\text{Im}(\tau)$.

As described in sections \ref{sec:mswcft} and \ref{n4ym} the (MSW) CFT and the ${\cal N}=4$ $U(r)$ Super-Yang-Mills partition functions should behave covariantly under modular transformations of the SL$(2,\IZ)$ acting on $\tau$. Thus, the modular completion outlined above will effect the generating functions $f^{(2)}_{\mu,J}$ through their relation to the indefinite theta-function $\Theta^{J,J'}_{\Lambda,\mu}$, which needs a modular completion to transform covariantly under modular transformations, i.e. \be\Theta^{J,J'}_{\Lambda,\mu} \mapsto \widehat{\Theta}^{J,J'}_{\Lambda, \mu} \ee and consequently $f^{(2)}_{\mu,J}$ is replaced by $\hat{f}^{(2)}_{\mu,J}$. Due to eq.~(\ref{mock}) the counting function of BPS invariants $\hat{f}^{(2)}_{\mu,J}$ and thus the elliptic genus $Z^{(2)}_P$ are going to suffer from a holomorphic anomaly, to which we turn next.

\section{Wall-crossing and mock modularity}\label{sec:wcmm}

In this section we derive an anomaly equation for two M5-branes wound on a rigid surface/divisor $P$ with $b_2^+(P)=1$, inside a Calabi-Yau manifold $X$. We begin by reviewing D4-D2-D0 bound-states in the type IIA picture and their wall-crossing in the context of the Kontsevich-Soibelman formula. Then we proceed by deriving a generating function for rank two sheaves from the Kontsevich-Soibelman formula which is equivalent to G\" ottsche's formula \cite{Gottsche:1999}. This generating function is an indefinite theta-function, which fails to be modular. As a next step we apply ideas of Zwegers to remedy this failure of modularity by introducing a non-holomorphic completion. This leads to a holomorphic anomaly equation of the elliptic genus of two M5-branes that we prove for rigid divisors $P$.

\subsection{D4-D2-D0 wall-crossing}\label{sec:wc}
In the following we take on the equivalent type IIA point of view, adapting the discussion of refs. \cite{Diaconescu:2007bf,Manschot:2010xp,Manschot:2010nc} to describe the relation to the Kontsevich-Soibelman wall-crossing formula \cite{KS:2008}. We restrict our attention to the D4-D2-D0 system on the complex surface $P$ and work in the large volume limit with vanishing $B$-field.

Let us recall that a generic charge vector with D4-brane charge $r$ is given by (see appendix \ref{Dbound} for details)
\begin{equation} \label{chv}
    \Gamma = (Q_6,Q_4,Q_2,Q_0) = r \left(0,\, [P],\, i_* F(\cE),\, \frac{\chi(P)}{24} + \int_P \half F(\cE)^2 - \Delta(\cE) \right),
\end{equation}
where $\cE$ is a sheaf on the divisor $P$. Further, we define
\begin{equation}
    \Delta(\cE)=\frac{1}{r(\cE)} \left(c_2(\cE) -\frac{r(\cE)-1}{2 r(\cE)}c_1(\cE)^2\right)\, , \quad \mu(\cE)=\frac{c_1(\cE)}{r(\cE)}\, ,
    \quad F(\cE) = \mu(\cE) + \frac{[P]}{2}.
\end{equation}
We recall that in the large volume regime the notion of D-brane stability is equivalent to $\mu$-stability, see \cite{Diaconescu:2007bf} and appendix~\ref{Dbound}. Given a choice of $J\in\cC(P)$, a sheaf $\cE$ is called $\mu$-semi-stable if for every sub-sheaf $\cE'$
\begin{equation}\label{stability}
    \mu(\cE')\cdot J \le \mu({\cE})\cdot J.
    \end{equation}
Moreover, a wall of marginal stability is a co-dimension one subspace of the K\"ahler cone $\cC(P)$ where the following condition is satisfied
\begin{equation} \label{muwall}
    (\mu(\cE_1)-\mu(\cE_2))\cdot J=0,
\end{equation}
but is non-zero away from the wall. Across such a wall of marginal stability the configuration (\ref{chv}) splits into two configurations with charge vectors
\begin{eqnarray}
    \Gamma_1 & = & r_1 \left(0,\, [P],\, i_* F_1,\, \frac{\chi(P)}{24} + \int_P \half F_1^2
              - \Delta(\mathcal{E}_1)\right), \nonumber \\
    \Gamma_2 & = & r_2 \left(0,\, [P],\, i_* F_2,\, \frac{\chi(P)}{24} + \int_P \half F_2^2
              - \Delta(\mathcal{E}_2)\right), \label{chv12}
\end{eqnarray}
where $r_i = \text{rk}(\mathcal{E}_i)$ and $\mu_i=\mu(\cE_i)$. By making use of the identity
\begin{eqnarray}\label{c1split}
      r \Delta & = & r_1 \Delta_1 + r_2 \Delta_2 + \frac{r_1 r_2}{2r} \left(\frac{c_1(\mathcal{E}_1)}{r_1} - \frac{c_1(\mathcal{E}_2)}{r_2}\right)^2,
\end{eqnarray}
one can show that $\Gamma = \Gamma_1 + \Gamma_2$. Therefore, charge-vectors as defined in (\ref{chv}) form a vector-space which will be
essential for the application of the Kontsevich-Soibelman formula.

Before we proceed, let us note, that the BPS numbers and the Euler numbers of the moduli space of sheaves are related as follows. Denote by $\mathcal{M}_{J}(\Gamma)$ the moduli space of semi-stable sheaves characterized by $\Gamma$. Its dimension reads \cite{Maruyama:1977}
\begin{equation}
    \textrm{dim}_{\mathbbm{C}} \mathcal{M}_J(\Gamma)= 2 r^2 -r^2 \chi(\mathcal{O}_P)+1.
\end{equation}
The relation between BPS invariants and the Euler numbers of the moduli spaces $\mathcal{M}_J(\Gamma)$
is then given by \cite{Diaconescu:2007bf}
\begin{equation}
    \Omega(\Gamma,J)= (-1)^{\textrm{dim}_{\mathbbm{C}} \mathcal{M}_J(\Gamma)} \chi(\mathcal{M}(\Gamma),J)\, .
\end{equation}
Moreover, for the system of charges we have specified to, the symplectic pairing of charges simplifies to \cite{Diaconescu:2007bf}
\begin{equation}\label{eq:chargeproduct}
    \langle \Gamma_1,\Gamma_2\rangle = r_1 r_2 (\mu_2-\mu_1) \cdot \left[ P\right].
\end{equation}

The holomorphic function $f^{(r)}_{\mu,J}(\tau)$ appearing in eq.~(\ref{thetadecomp}) can now be identified with the generating function of BPS invariants of moduli spaces of semi-stable sheaves. Its wall crossing will be described in the following.

\subsubsection*{\it Kontsevich-Soibelman wall-crossing formula}
Kontsevich and Soibelman \cite{KS:2008} have proposed a formula which determines the jumping behavior of BPS-invariants $\Omega(\Gamma; J)$ across walls of marginal stability.  The wall-crossing formula is given in terms of a Lie algebra defined by generators $e_{\Gamma}$ and a basic commutation relation
\begin{equation}
  \left[e_{\Gamma_1}, e_{\Gamma_2}\right] = (-1)^{\langle \Gamma_1, \Gamma_2 \rangle} \langle \Gamma_1, \Gamma_2 \rangle e_{\Gamma_1 + \Gamma_2}.
\end{equation}

For every charge $\Gamma$ an element $U_{\Gamma}$ of the Lie group can be defined by
\begin{equation}
  U_{\Gamma} = \textrm{exp}\left(-\sum_{n \geq 1} \frac{e_{n\Gamma}}{n^2} \right).
\end{equation}

The Kontsevich-Soibelman wall-crossing formula states that across a wall of marginal stability the following formula holds
\begin{equation} \label{kswc}
  \prod_{\Gamma: Z(\Gamma; J) \in V}^{\curvearrowright} U_{\Gamma}^{\Omega(\Gamma; J_+)}
  = \prod_{\Gamma: Z(\Gamma; J) \in V}^{\curvearrowright} U_{\Gamma}^{\Omega(\Gamma; J_-)},
\end{equation}
where $J_+$ and $J_-$ denote K\"ahler classes on the two sides of the wall. Further, $V$ is a region in $\IR^2$ bounded by two rays starting at the origin and $\curvearrowright$ denotes a clockwise ordering of the factors in the product with respect to the phase of the central charges $Z(\Gamma; J)$, that are defined in eq.~(\ref{eq:centralcharge}).

Restricting to the case $r=2$ and $r_1 = r_2 = 1$, (\ref{kswc}) can be truncated to
\begin{equation}
    \prod_{Q_{0,1}} U_{\Gamma_1}^{\Omega(\Gamma_1)} \prod_{Q_0} U_{\Gamma}^{\Omega(\Gamma; J_+)}
    \prod_{Q_{0,2}} U_{\Gamma_2}^{\Omega(\Gamma_2)}
    =
    \prod_{Q_{0,2}} U_{\Gamma_2}^{\Omega(\Gamma_2)} \prod_{Q_0} U_{\Gamma}^{\Omega(\Gamma; J_-)}
    \prod_{Q_{0,1}} U_{\Gamma_1}^{\Omega(\Gamma_1)},
\end{equation}
where $Q_0$ is the D0-brane charge of $\Gamma$ and the $Q_{0,i}$ are the D0-brane charges belonging to $\Gamma_i$, respectively.
The above formula has been derived by setting all Lie algebra elements with D4-brane charge greater than two to zero.
Therefore, the element $e_{\Gamma}$ is central, using the Baker-Campbell-Hausdorff formula $e^X e^Y = e^Y e^{[X,Y]} e^X$ and the fact that the symplectic product is independent of the D0-brane charge,
one finds the following change of BPS numbers across a wall of marginal stability \cite{Manschot:2010xp, Gaiotto:2008cd}
\begin{equation} \label{pwc}
    \Delta \Omega(\Gamma) = (-1)^{\langle \Gamma_1, \Gamma_2 \rangle - 1} \langle \Gamma_1, \Gamma_2 \rangle
                            \sum _{Q_{0,1}+Q_{0,2}=Q_0} \Omega(\Gamma_1)\, \Omega(\Gamma_2).
\end{equation}
Moreover, one can deduce that the rank one degeneracies $\Omega(\Gamma_1)$ and $\Omega(\Gamma_2)$ do not depend on the modulus $J$.

\subsection{Relation of KS to G\"ottsche's wall-crossing formula}\label{sec:KSGwallcrossing}
G\"ottsche has found a wall-crossing formula for the Euler numbers of moduli spaces of rank two sheaves in terms of an indefinite theta-function in ref.~\cite{Gottsche:1999}.
In this section we want to derive a modified version of this formula from the Kontsevich-Soibelman wall-crossing formula associated to D4-D2-D0 bound-states with D4-brane charge equal to two.

We use the short notation $\Gamma=(r,\mu,\Delta)$ to denote a rank $r$ sheaf with the specified Chern classes that is associated to the D4-D2-D0 states. For rank one sheaves the generating function has no chamber dependence and we have already seen that it is given by (\ref{frigid}).
Following the discussion of our last section, higher rank sheaves do exhibit wall-crossing phenomena and therefore do depend on the chamber in
moduli space, i.e.~on $J\in\cC(P)$.  

Our aim now is to determine the generating function of the D4-D2-D0 system using the primitive wall-crossing formula derived from the KS wall-crossing formula. From now on we restrict our attention to rank two sheaves $\mathcal{E}$. They can split across walls of marginal stability into rank one sheaves $\mathcal{E}_1$
and $\mathcal{E}_2$ as outlined in section \ref{sec:wc}. Using relation (\ref{c1split}) we can write
\begin{equation} \label{Deltarec}
    d = d_1 + d_2 + \xi\cdot\xi,
\end{equation}
where $\xi= \mu_1-\mu_2$ and $d=2\Delta$. Further, a wall is given by (\ref{muwall}), i.e.~the set of walls given a split of charges $\xi$ reads
\begin{equation}
    W^{\xi} = \left\{J \in \cC(P)\,|\, \xi\cdot J = 0\right\}.
\end{equation}
Now, consider a single wall $J_W\in W^{\xi}$ determined by a set of vectors $\xi \in \Lambda + \mu$. Let $J_+$ approach $J_W$ infinitesimally close from one side and $J_-$ infinitesimally close from the other side. Thus, in our context the primitive wall-crossing formula (\ref{pwc}) becomes
\begin{equation} \label{cswc}
    \bar{\Omega}(\Gamma;J_+) - \bar{\Omega}(\Gamma;J_-)
    = \sum_{Q_{0,1}+Q_{0,2}=Q_0} (-1)^{2\xi\cdot[P]}\, 2\,  (\xi\cdot[P])\, \Omega(\Gamma_1)\, \Omega(\Gamma_2),
\end{equation}
where we  have used the identity \eqref{eq:chargeproduct}. Note, that $Q_{0,i}$ and $Q_0$ are determined in terms of $\Gamma$ and $\Gamma_i$ through (\ref{chv}) and (\ref{chv12}). Now, we can sum over the D0-brane charges to obtain a generating series. This yields 
\begin{eqnarray}
    & ~ & \sum_{d\,\geq\,0} (\bar{\Omega}(\Gamma;J_+) - \bar{\Omega}(\Gamma;J_-)) q^{d - \frac{\chi(P)}{12}} \nonumber \\
    ~ ~ & = & \sum_{d_1,d_2\,\geq\,0,\,\xi} (-1)^{2 \xi\cdot [P]}\,  (\xi\cdot [P])\, \Omega(\Gamma_1) \Omega(\Gamma_2)
            q^{d_1 + d_2 + \xi^2 - \frac{2\chi(P)}{24}} \nonumber \\
    ~ ~ & = & (-1)^{2 \mu\cdot[P]-1}\frac{\vartheta_{\Lambda^\perp}(\tau)^2}{\eta(\tau)^{2\chi(P)}}\,\sum_{\xi} (\xi\cdot[P])\, q^{\xi^2},
\end{eqnarray}
where for the first equality use has been made of the identities (\ref{Deltarec}, \ref{cswc}), and for the second equality
the identity (\ref{frigid}) has been used. The last line can be rewritten as
\begin{equation}
    (-1)^{2 \mu\cdot[P]-1} \frac{1}{2}\frac{\vartheta_{\Lambda^\perp}(\tau)^2}{\eta(\tau)^{2 \chi(P)}} \textrm{Coeff}_{2 \pi i y} (\Theta^{J_+,J_-}_{\Lambda,\mu}( \tau, [P] y)),
\end{equation}
where we have introduced the indefinite theta-function
\begin{equation} \label{eq:indefinite}
    \Theta^{J,J'}_{\Lambda,\mu}(\tau,x) :=\frac{1}{2} \sum_{\xi \in \Lambda + \mu} (\sgn\langle J,\xi\rangle-\sgn\langle J', \xi\rangle)\, e^{2 \pi i \langle \xi, x\rangle}\, q^{Q(\xi)},
\end{equation}
with the inner product\footnote{Note, that this is not the symplectic product of D-brane charges defined before.} defined by $\langle x, y\rangle = 2 d_{AB} x^A y^B$ and the quadratic form $Q(\xi)=\frac{1}{2} \langle \xi, \xi \rangle$. As these theta-functions obey the cocycle condition \cite{Gottsche:1998}
\begin{equation}
    \Theta^{F,G}_{\Lambda,\mu} + \Theta^{G,H}_{\Lambda,\mu} = \Theta^{F,H}_{\Lambda,\mu},
\end{equation}
we finally arrive at the beautiful relation between the BPS numbers in an arbitrary chamber $J$ and those in a chamber $J'$ first found by G\"ottsche in the case $\Lambda=H^2(P,\IZ)$:
\be\label{eq:thm41}
f^{(2)}_{\mu,J'}(\tau)-f^{(2)}_{\mu,J}(\tau)=\frac{1}{2}\frac{\vartheta_{\Lambda^\perp}(\tau)^2}{\eta^{2\chi(P)}(\tau)}\,\text{Coeff}_{2\pi i y}(\Theta^{J,J'}_{\Lambda,\mu}( \tau,[P] y)).
\ee

\subsection{Holomorphic anomaly at rank two}
In this subsection we discuss the appearance of a holomorphic anomaly at rank two and give a proof of it by combing our previous results with results of Zwegers \cite{Zwegers:2002}.

\subsubsection{Elliptic genus at rank two and modularity}
An important datum in eq.~(\ref{eq:thm41}) is the choice of chambers $J,J'\in\cC(P)$, which are any points in the K\"ahler cone of $P$. As a consequence, the indefinite theta-series does not transform well under SL$(2,\IZ)$ in general. However, from the discussion of sect.~\ref{sec:mswcft} we expect, that the generating series $f^{(r)}_{\mu,J}(\tau)$ transforms with weight $-\frac{r(\Lambda)+2}{2}$ in a vector-representation under the full modular group, where $r(\Lambda)$ denotes the rank of the lattice $\Lambda$. Hence, there is a need to restore modularity. The idea is as follows.

Following Zwegers \cite{Zwegers:2002}, it turns out that the indefinite theta-function can be made modular at the cost of losing its holomorphicity. From the definition (\ref{eq:indefinite}) Zwegers smoothes out the sign-functions and introduces a modified function as
\be\label{eq:mocktheta}
\widehat{\Theta}^{c,c'}_{\Lambda,\mu}(\tau,x)=\frac{1}{2}\sum_{\xi\,\in\,\Lambda+\mu}\left(\!\!E\!\left(\frac{\langle c, \xi+\frac{\text{Im}\,(x)}{\tau_2}\rangle\sqrt{\tau_2}}{\sqrt{-Q(c)}}\right)-E\!\left(\frac{\langle c',\xi+\frac{\text{Im}\,(x)}{\tau_2}\rangle\sqrt{\tau_2}}{\sqrt{-Q(c')}}\right)\!\!\right)e^{2\pi i \langle \xi, x\rangle} q^{Q(\xi)},
\ee
where $E$ denotes the incomplete error function
\be
E(x)=2\int_0^x e^{-\pi u^2}du.
\ee
Note, that if $c$ or $c'$ lie on the boundary of the K\"ahler cone, one does not have to smooth out the sign-function. Zwegers shows, that the non-holomorphic function $\widehat{\Theta}^{c,c'}_{\Lambda,\mu}(\tau,x)$ satisfies the correct transformation properties of a Jacobi form of weight $\frac{1}{2}r(\Lambda)$. Due to the non-holomorphic pieces it contains mock modular forms, that we want to identify in the following. In order to separate the holomorphic part of $\widehat{\Theta}^{c,c'}_{\Lambda,\mu}(\tau,x)$ from its shadow we recall the following property of the incomplete error function
\be
E(x)=\text{sgn}(x)(1-\beta_{\frac{1}{2}}(x^2)),
\ee
which enables us to split up $\widehat{\Theta}^{c,c'}_{\Lambda,\mu}(\tau,x)$ into pieces. Here, $\beta_{k}$ is defined by
\begin{equation}
\beta_k(t)=\int_t^{\infty} u^{-k} e^{-\pi u} du.
\end{equation}
Hence, one can write eq.~\eqref{eq:mocktheta} as
\be
\widehat{\Theta}^{c,c'}_{\Lambda,\mu}(\tau,x)=\Theta^{c,c'}_{\Lambda,\mu}(\tau,x)-\Phi^{c}_{\mu}(\tau,x)+\Phi^{c'}_{\mu}(\tau,x),
\ee
with
\be
\Phi^{c}_{\mu}(\tau,x)=\frac{1}{2}\sum_{\xi\,\in\,\Lambda+\mu}\left[\text{sgn}\langle \xi, c\rangle-E\left(\frac{\langle c , \xi+\frac{\text{Im}\,(x)}{\tau_2}\rangle \sqrt{\tau_2}}{\sqrt{-Q(c)}}\right)\right]e^{2\pi i \langle \xi,  x\rangle} q^{Q(\xi)}.
\ee
If $c$ belongs to ${\cal C}(P)\cap \mathbb{Q}^{r(\Lambda)}$, we may write
\be
\Phi^{c}_{\mu}(\tau,x)=R(\tau,x)\theta(\tau,x),
\ee
where we decomposed the lattice sum into contributions along the direction of $c$ and perpendicular to $c$ given by $R$ and $\theta$, respectively. Hence, $\theta$ is a usual theta-series associated to the quadratic form $Q|\langle c\rangle^\perp$, i.e.~of weight $(r(\Lambda)-1)/2$. $R$ is the part which carries the non-holomorphicity. It transforms with a weight $\frac{1}{2}$ factor and therefore $\text{Coeff}_{2\pi iy}(R(\tau,[P]y))$ is of weight $\frac{3}{2}$. Following the general idea of Zagier \cite{Zagier:2007} that we recapitulate in appendix \ref{sec:mock}, we should encounter the $\beta_{\frac{3}{2}}$ function in the $2\pi iy$-coefficient of $\Phi$. Indeed one can prove the following identity
\be
\text{Coeff}_{2\pi iy} \Phi^{c}_{\mu}(\tau,[P]y)=-\frac{1}{4\pi}\frac{\langle c,[P]\rangle}{\langle c,c\rangle}\sum_{\xi\,\in\,\Lambda+\mu}|\langle c,\xi\rangle|\, \beta_{\frac{3}{2}}\left(\frac{\tau_2 \langle c,\xi\rangle^2}{-Q(c)}\right)q^{Q(\xi)}.
\ee
Taking the derivative with respect to $\bar \tau$ in order to obtain the shadow we arrive at the following final expression
\be\label{aniholomder}
\partial_{\bar \tau} \text{Coeff}_{2\pi i y} \Phi^{c}_\mu (\tau, [P] y)=-\frac{\tau_2^{-\frac{3}{2}}}{8 \pi i} \frac{c\cdot[P]}{\sqrt{-c^2}}\,(-1)^{4{\mu}^2}\, \theta^{(2)}_{\mu-\frac{[P]}{2},c}(\tau,0),
\ee
where we define the Siegel-Narain theta-function $\theta^{(r)}_{\mu,c}(\tau,z)$ as in eq.~(\ref{SNTheta}). For more details on the transformation properties of the indefinite theta-functions we refer the reader to appendix \ref{sec:mock}.

Now, these results can be used to compute the elliptic genus for two M5-branes wrapping the divisor $P$. Consider
\be
f^{(2)}_{\mu,J}(\tau)=f_{\mu,J'}(\tau)-\frac{1}{2}\frac{\vartheta_{\Lambda^\perp}(\tau)^2}{\eta^{2\chi(P)}}\,\text{Coeff}_{2\pi i y}\Theta^{J,J'}_{\Lambda,\mu}(\tau,[P]y),
\ee
where $f_{\mu,J'}(\tau)$ is a holomorphic ambiguity given by the generating series in a reference chamber $J'$, which we choose to lie at the boundary of the K\"ahler cone $J'\,\in\,\p\cC(P)$. In explicit computations it may be possible to choose $J'$ such that the BPS numbers vanish. In general, however, such a vanishing chamber might not always exist, but since $J'$ is at the boundary of the K\"ahler cone, $f_{\mu,J'}(\tau)$ has no influence on the modular transformation properties, nor on the holomorphic anomaly. We write the full M5-brane elliptic genus as
\be
Z^{(2)}_{P}(\tau,z)=\sum_{\mu\,\in\,\Lambda^*/\Lambda}\hat{f}^{(2)}_{\mu,J}(\tau)\theta^{(2)}_{\mu,J}(\tau,z),
\ee
where $\hat{f}^{(2)}_{\mu,J}$ denotes the modular completion as outlined above. We can show using Zwegers' results \cite{Zwegers:2002}, that the M5-brane elliptic genus transforms like a Jacobi form of bi-weight $(-\frac{3}{2},\frac{1}{2})$. Again, we refer the reader to appendix \ref{sec:mock} for further details.

\subsubsection{Proof of holomorphic anomaly at rank two}\label{sec:holanomalyrantwo}
Now, we are in position to prove the holomorphic anomaly at rank two for general surfaces $P$ with $b_2^+(P)=1$.  The holomorphic anomaly takes the following form
\be\label{eq:provenholan}
{\cal D}_2 Z^{(2)}_{P}(\tau,z)=\tau_2^{-3/2}\frac{1}{16\pi i}\frac{J\cdot[P]}{\sqrt{-J^2}}\left(Z^{(1)}_{P}(\tau,z)\right)^2,
\ee
where the derivative ${\cal D}_k$ is given as
\be
{\cal D}_k=\p_{\bar\tau}+\frac{i}{4\pi k}\p^2_{z_+},
\ee
and $z_+$ refers to the projection of $z$ along a direction $J\in{\cal C}(P)$. For the proof, ${\cal D}_2 Z^{(2)}_{P}$ can be computed explicitly. Using \eqref{aniholomder} we obtain directly
\be
{\cal D}_2 Z^{(2)}_{P}(\tau,z)=\tau_2^{-3/2}\frac{1}{16\pi i}\frac{J\cdot [P]}{\sqrt{-J^2}}\frac{\vartheta_{\Lambda^\perp}(\tau)^2}{\eta(\tau)^{2\chi}}\,\sum_{\mu\,\in\,\Lambda^*/\Lambda}(-1)^{4\mu^2}\theta^{(2)}_{\mu-\frac{[P]}{2},J}(\tau,0)\theta_{\mu,J}^{(2)}(\tau,z).
\ee
Since the following identity among the theta-functions $\theta_{\mu,J}$ holds
\be\label{theta-identity1}
\left(\theta_{0,J}^{(1)}(\tau,z)\right)^2=\sum_{\mu\,\in\,\Lambda^*/\Lambda}(-1)^{4\mu^2}\theta^{(2)}_{\mu-\frac{[P]}{2},J}(\tau,0)\theta_{\mu,J}^{(2)}(\tau,z),
\ee
we have proven the holomorphic anomaly equation at rank two for general surfaces $P$.

\section{Applications and extensions}
In the following we want to apply the previous results to several selected examples. Before doing so, we explain two mathematical facts which will help to fix the ambiguity 
$f_{\mu,J'}(\tau)$, which are the blow-up formula and the vanishing lemma. After discussing the examples, we turn our attention to a possible extension to higher rank. This leads us to speculations about mock modularity of higher depth and wall-crossing having its origin in a meromorphic Jacobi form.

For the modular forms used in this section, we refer the reader to appendix \ref{sec:mock}.

\subsection{Blow-up formulae and vanishing chambers}
There is a universal relation between the generating functions of stable sheaves on a surface $P$ and on its blow-up $\tilde P$ \cite{Vafa:1994tf, Yoshioka:1995, Yoshioka:1996, Li:1999, Gottsche:1999}. Let $P$ be a smooth projective surface and $\pi:\tilde{P}\rightarrow P$ the blow-up at a non-singular point with $E$ the exceptional divisor of $\pi$. Let $J\in\cC(P)$, $r$ and $\mu$ such that gcd$(r,r\mu\cdot J)=1$. Then, the generating series $f^{(r)}_{\mu,J}(\tau;P)$ and $f^{(r)}_{\mu,J}(\tau;\tilde P)$ are related by the blow-up formula
\be\label{blowupformula}
f^{(r)}_{\pi^*(\mu)-\frac{k}{r} E,\pi^*(J)}(\tau;\tilde P)=B_{r,k}(\tau)f^{(r)}_{\mu,J}(\tau;P),
\ee
with $B_{r,k}$ given by
\be
B_{r,k}(\tau)=\frac{1}{\eta^r(\tau)}\sum_{a\,\in\,\IZ^{r-1}+\frac{k}{r}}q^{\sum_{i\leq j}a_i a_j}.
\ee

The second fact states that for a class of semi-stable sheaves on certain surfaces the moduli space of the sheaves is empty. We refer to this fact as the vanishing lemma \cite{Gottsche:1999}. For this let $P$ be a rational ruled surface $\pi:P\rightarrow\IP^1$ and $J$ be the pullback of the class of a fiber of $\pi$. Picking a Chern class $\mu$ with $r\mu\cdot J$ odd, we have
\be
{\cal M}((r,\mu,\Delta),J)=\emptyset
\ee
for all $d$ and $r\geq 2$.

\subsection{Applications to surfaces with $b_2^+=1$}\label{sec:surfaceswithb+=1}
The surfaces we are going to consider are $\mathbb{P}^2$, the Hirzebruch surfaces $\mathbb{F}_0$ and $\mathbb{F}_1$, the del Pezzo surfaces $\mathcal{B}_8$ and $\mathcal{B}_9$.

\subsubsection*{\it Projective plane $\mathbb{P}^2$}
The projective plane $\mathbb{P}^2$ has been discussed quite exhaustively in the literature. The rank one result was obtained by G\"ottsche \cite{Gottsche:1990}
\be
Z^{(1)}_{\mathbb{P}^2}=\frac{\vartheta_1(-\bar\tau,-z)}{\eta^{3}(\tau)}.
\ee
The generating functions of the moduli space of rank two sheaves or $SO(3)$ instantons of Super-Yang-Mills theory on $\mathbb{P}^2$ were written down by \cite{Yoshioka:1994,Yoshioka:1995,Vafa:1994tf} and are given by
\be\label{expansionforp2}
\begin{split}
f_0(\tau)&=\sum_{n=0}^\infty\chi(\cM((2,0,n),J))q^{n-\frac{1}{4}}=\frac{3h_0(\tau)}{\eta^6(\tau)},\\
f_1(\tau)&=\sum_{n=0}^\infty\chi(\cM((2,1,n),J))q^{n-\frac{1}{2}}=\frac{3h_1(\tau)}{\eta^6(\tau)}.\\
\end{split}
\ee
Here, $h_j(\tau)$ are mock modular forms given by summing over Hurwitz class numbers $H(n)$
\be
h_j(\tau)=\sum_{n=0}^\infty H(4n+3j)q^{n+\frac{3j}{4}},\qquad (j=0,1).
\ee
Their modular completion is denoted by $\hat{h}_j(\tau)$, where the shadows are given by $\vartheta_{3-j}(2\tau)$ \cite{Zagier:1975}. Explicitly, we have
\be
\partial_{\bar\tau}\hat{h}_j(\tau)=\frac{\tau_2^{-\frac{3}{2}}}{16\pi i}\vartheta_{3-j}(-2\bar\tau).
\ee
Note, that these results are valid for all K\"ahler classes $J\in H^2(\IP^2,\IZ)$ as there is no wall crossing in the K\"ahler moduli space of $\IP^2$. This leads directly to the following elliptic genus of two M5-branes wrapping the $\mathbb{P}^2$ divisor
\be
Z^{(2)}_{\mathbb{P}^2}(\tau,z)=\hat{f}_0(\tau)\vartheta_2(-2\bar\tau,-2z)-\hat{f}_1(\tau)\vartheta_3(-2\bar\tau,-2z).
\ee
Denoting by ${\cal D}_2=\p_{\bar\tau}+\frac{i}{8\pi}\p_{z}^2$ one finds the expected holomorphic anomaly equation at rank two, given by\footnote{This result has already been derived in \cite{Bringmann:2010sd}.}
\be
{\cal D}_2\, Z^{(2)}_{\mathbb{P}^2}(\tau,z)=-\frac{3}{16\pi i}\tau_2^{-\frac{3}{2}}\left(Z^{(1)}_{\mathbb{P}^2}(\tau,z)\right)^2,
\ee
which can be derived directly from the simple fact that
\be
\vartheta_1(\tau,z)^2=\vartheta_2(2\tau)\vartheta_3(2\tau,2z)-\vartheta_3(2\tau)\vartheta_2(2\tau,2z).
\ee

Further note, that the $q$-expansion of $f_0$, eq.~(\ref{expansionforp2}), has non-integer coefficients. It was explained in \cite{Manschot:2010xp} that this is due to the fact that the generating series involves the fractional BPS invariants $\bar\Omega(\Gamma)$, which we encountered before.

\subsubsection*{\it Hirzebruch surface $\mathbb{F}_0$}
Our next example is the Hirzebruch surface $P=\mathbb{F}_0$. We denote by $F$ and $B$ the fiber and the base $\IP^1$'s respectively. For an embedding into a Calabi-Yau manifold one may consult app.~\ref{toricdata}. Let us choose $J=F+B$, $J'=B$ and Chern class $\mu=F/2$. The choice $\mu=B/2$ can be treated analogously and leads to the same results. The other sectors corresponding to $\mu=0$ and $\mu=(F+B)/2$ require a knowledge of the holomorphic ambiguity at the boundary and will not be treated here. One obtains
\be
\begin{split}
f^{(2)}_{\mu,F+B}(\tau)&=\frac{1}{2\eta^8(\tau)}\text{Coeff}_{2\pi iy}(\Theta^{F+B,B}_{\Lambda,\mu}(\tau,[P] y))\\ &=q^{-\frac{1}{3}}\left(2q+22q^2+146q^3+742q^4+\dots\right),
\end{split}
\ee
where we denote by $\mu$ either $B/2$ or $F/2$. This exactly reproduces the numbers obtained in \cite{Kool}.

We want to compute the shadow of the completion given by adding $\Phi^{F+B}_{\mu}$ and $\Phi^{B}_{\mu}$ to the indefinite theta-series $\Theta^{F+B,B}_{\Lambda,\mu}$. Since $B$ is chosen at the boundary, $\Phi^{B}_{\mu}$ vanishes for $\mu=F/2,B/2$. The only relevant contribution has a shadow proportional to $\vartheta_2(\tau)$. Precisely, we obtain
\be
\partial_{\bar\tau}f_{\mu,F+B}^{(2)}(\tau)=-\tau_2^{-3/2}\frac{1}{4\pi i\sqrt{2}}\,\frac{\overline{\vartheta_2(\tau)}\vartheta_2(\tau)}{\eta^8(\tau)}\qquad (\mu=\frac{F}{2},\frac{B}{2}).
\ee

\subsubsection*{\it Hirzebruch surface $\mathbb{F}_1$}
The next example is the Hirzebruch surface $\mathbb{F}_1$, which is a blow-up of $\mathbb{P}^2$. Again we denote by $F$ and $B$ the fiber and base $\IP^1$'s. The $\mathbb{P}^2$ hyperplane is given by the pullback of $F+B$ and $B$ is the exceptional divisor. This example is particularly nice, since we can check our results against the blow-up formula \eqref{blowupformula} or use the results known from $\mathbb{P}^2$ to write generating functions in sectors which are not accessible through the vanishing lemma. Notice, that the holomorphic expansions have been already discussed in ref.~\cite{Manschot:2010nc}. From the general discussion one sees that there are four different choices for the Chern class $\mu\in\{\frac{B}{2},\frac{F+B}{2},\frac{F}{2},0 \}$.

First, we choose $J=F+B$, $J'=F$ and Chern class $\mu=B/2$. We then obtain
\be\label{eqF+BF1}
\begin{split}
f^{(2)}_{\mu,F+B}(\tau)&=\frac{1}{2\eta^8(\tau)}\text{Coeff}_{2\pi iy}(\Theta^{F+B,F}_{\Lambda,B}(\tau,[P] y))\\ &=q^{-\frac{1}{12}}\left(-\frac{1}{2} - q + \frac{15}{2} q^2 + 91 q^3 + 558 q^4+\dots\right).
\end{split}
\ee
A check of this result against the blow-up formula \eqref{blowupformula} applied to $\mathbb{P}^2$ yields
\be
\frac{3h_0(\tau)}{\eta^6(\tau)}\frac{\vartheta_2(2\tau)}{\eta^2(\tau)}=q^{-\frac{1}{12}}\left(-\frac{1}{2} - q + \frac{15}{2} q^2 + 91 q^3 + 558 q^4+\dots\right)=f^{(2)}_{\mu,F+B}(\tau).
\ee
Further, we calculate the shadow by differentiating $\hat{f}^{(2)}$ with respect to $\bar\tau$
\be
\partial_{\bar\tau}\hat{f}^{(2)}_{\mu,F+B}(\tau)=\frac{3}{16\pi i}\tau_2^{-3/2}\frac{\overline{\vartheta_3(2\tau)}\vartheta_2(2\tau)}{\eta^8(\tau)},
\ee
which also is in accord with the blow-up formula. Note, that \eqref{eqF+BF1} has half-integer expansion coefficients, since $J=B+F$ lies on a wall for the Chern class $\mu=B/2$.\\

As a second case we choose $J=F+B$, $J'=F$ and Chern class $\mu=(F+B)/2$ and obtain
\be
\begin{split}
f^{(2)}_{\mu,F+B}(\tau)&=\frac{1}{2\eta^8(\tau)}\text{Coeff}_{2\pi iy}(\Theta^{F+B,F}_{\Lambda,F+B}(\tau,[P]y))\\ &=q^{-\frac{7}{12}}\left(q + 13 q^2 + 93 q^3 + 496 q^4+\dots\right),
\end{split}
\ee
which we again can check against the blow-up formula \eqref{blowupformula} for $\IP^2$
\be
\frac{3h_1(\tau)}{\eta^6(\tau)}\frac{\vartheta_3(2\tau)}{\eta^2(\tau)}=q^{-\frac{7}{12}}\left(q + 13 q^2 + 93 q^3 + 496 q^4+\dots\right)=f^{(2)}_{\mu,F+B}(\tau).
\ee
Calculating the shadow yields
\be
\partial_{\bar\tau}\hat{f}^{(2)}_{\mu,F+B}(\tau)=\frac{3}{16\pi i}\tau_2^{-3/2}\frac{\overline{\vartheta_2(2\tau)}\vartheta_3(2\tau)}{\eta^8(\tau)},
\ee
which is also in accord with the blow-up formula.\\

The last two sectors $\mu=F/2,0$ are not accessible via the vanishing lemma. However, using a blow-down to $\IP^2$ we observe, that the above two cases reproduce correctly the two Chern classes in the cases of rank two sheaves on $\IP^2$. Using the blow-up formulas once more we finally arrive at
\be
\begin{split}
f^{(2)}_{(0,0),J}(\tau)&=\frac{3h_0(\tau)}{\eta^6(\tau)}\frac{\vartheta_3(2\tau)}{\eta^2(\tau)},\\
f^{(2)}_{(\frac{1}{2},0),J}(\tau)&=\frac{3h_1(\tau)}{\eta^6(\tau)}\frac{\vartheta_2(2\tau)}{\eta^2(\tau)},\\
f^{(2)}_{(0,\frac{1}{2}),J}(\tau)&=\frac{3h_0(\tau)}{\eta^6(\tau)}\frac{\vartheta_2(2\tau)}{\eta^2(\tau)},\\
f^{(2)}_{(\frac{1}{2},\frac{1}{2}),J}(\tau)&=\frac{3h_1(\tau)}{\eta^6(\tau)}\frac{\vartheta_3(2\tau)}{\eta^2(\tau)},
\end{split}
\ee
where $J=F+B$ and $\mu=(a,b)=aF+bB$. Note, that in the cases $f^{(2)}_{(0,0),J}$ and $f^{(2)}_{(0,\frac{1}{2}),J}$ the blow-up formula is not valid since we violate the gcd-condition, as $\pi_{*} \mu=0$ in these cases. However, for rank two sheaves on ${\mathbb F}_1$ the blow-up formula seems to work anyway, since the generating series using the blow-up procedure and the indefinite theta-function description coincide for the Chern class $\mu=(0,\frac{1}{2})$.

\subsubsection*{\it Del Pezzo surface $\mathcal{B}_8$}
As in \cite{Gaiotto:2006wm} we embed the surface $\mathcal{B}_8$ in a certain free $\IZ_5$ quotient\footnote{The only freely acting group actions for the quintic are a $\IZ_5^2$ and the above $\IZ_5$.} of the Fermat quintic $\tilde{X}=\{\sum_{i=1}^5 x_i^5=0\}$ in $\mathbb{P}^4$. The action of the group $G=\IZ_5$ on the projective coordinates of the ambient space is given by $x_i \sim \omega^i x_i $, where $\omega=e^{2\pi i /5}$. For the hyperplane section, denoted $P$, we observe that $P^3=1$, as for the Fermat quintic the five points of intersection of three hyperplanes $\{x_i=x_j=x_k=0\}$ are identified under the action of the group $G$. The Euler character of the hyperplane is given by $\chi(P) = 11$. It can be shown that the divisor $P$ is rigid and has $b_2^+=1$. We observe that  $H^2(P,\IZ) = \IZ \oplus (-E_8)$ as is explained in \cite{Gaiotto:2006wm}. The elliptic genus of a single M5-brane is then fixed by the modular weights
\ben
Z^{(1)}_{P}(\tau,z)= \frac{E_4(\tau)}{\eta^{11}(\tau)}\,\vartheta_1(-\overline{\tau},-z).
\een
The form of $Z^{(2)}_{P}$ can now be calculated as for $\IP^2$ and is given by
\ben
Z^{(2)}_{P}(\tau,z) \sim \frac{E_4(\tau)^2}{\eta(\tau)^{22}}\,(\hat{h}_0(\tau)\,{\vartheta}_2(-2\bar\tau,-2z)- \hat{h}_1(\tau)\,{\vartheta}_3(-2\bar\tau,-2z)).
\een
The holomorphic anomaly equation fulfilled by $Z^{(2)}_{P}(\tau,z)$ can be obtained as in the $\IP^2$ case
\ben
{\cal D}_2\, Z^{(2)}_{P}(\tau,z)\sim\frac{\tau_2^{-\frac{3}{2}}}{16\pi i}\left(Z^{(1)}_{P}(\tau,z)\right)^2.
\een

\subsubsection*{\it Del Pezzo surface $\mathcal{B}_9$, the $\frac{1}{2}$K3}
We end our examples by returning and commenting on $\frac{1}{2}$K3 or $\mathcal{B}_9$ which was the example of section (\ref{n4ym}), as M5-branes wrapping on it give rise to the multiple E-strings. The  $\mathcal{B}_9$ surface can be understood as a $\IP^2$ blown up at nine points (see appendix \ref{geometry} for details) or a rational elliptic surface. This case is interesting as one can map via T-duality along the elliptic fibration the computation of the modified elliptic genus to the computation of the partition function of topological string theory on the same surface \cite{Minahan:1998vr}. The middle dimensional cohomology lattice of $\mathcal{B}_9$ is given by $H^2(\mathcal{B}_9,\IZ)=\Gamma^{1,1}\oplus E_8$ and the Euler number can be computed to $\chi(\mathcal{B}_9)=12$. Modularity then fixes the form of the elliptic genus at rank one to
\ben
Z^{(1)}_{\mathcal{B}_9}(\tau,z)=\frac{E_4(\tau)}{\eta(\tau)^{12}}\,\theta^{(1)}_{0,J}(\tau,z),
\een
where $\theta^{(1)}_{0,J}(\tau,z)$ is the theta-function associated to the lattice $\Gamma^{1,1}$ with standard intersection form
\be
(-d_{AB})=\begin{pmatrix}
          0&1\\ 1&0
         \end{pmatrix}
.
\ee
Choosing the K\"ahler form $J=(R^{-2},1)^T$, where $(1,0)^T$ is the class of the elliptic fiber, one can show that
\be
\theta^{(1)}_{0,J}(\tau,0)\rightarrow \frac{R}{\sqrt{\tau_2}}\quad \text{as}\quad R\rightarrow\infty.
\ee
In this limit of small elliptic fiber one recovers the results of sect.~\ref{n4ym}. The factor $E_4(\tau)$ is precisely the theta-function of the $E_8$ lattice. The results obtained from the anomaly for higher wrappings of refs.~\cite{Minahan:1997ct,Minahan:1998vr} were proven mathematically for double wrapping in ref.~\cite{Yoshioka:1999}. In this analysis the Weyl group of the $E_8$ lattice was used to perform the theta-function decomposition.

\subsection{Extensions to higher rank and speculations}
In the following sections we want to discuss the extension of our results to higher rank. Partial results for rank three can be found already in the literature \cite{Kool,Manschot:2010nc,Weist:2009,Yoshioka:2008,Klyachko}. Thereafter, we discuss a possible generalization of mock modularity and speculate about a contour description which stems from a relation to a meromorphic Jacobi form.

\subsubsection{Higher rank anomaly and mock modularity of higher depth}
We want to focus on the holomorphic anomaly equation at general rank as conjectured in \cite{Minahan:1998vr}. We recall that its form is given by
\be\label{eq:anomalygeneral}
{\cal D}_r Z^{(r)}_{P} (\tau, z) \sim  \sum_{n=1}^{r-1} n(r-n) Z^{(n)}_{P}(\tau, z) Z^{(r-n)}_{P}(\tau, z),
\ee
where $Z^{(r)}_{P}(\tau, z)$ can be decomposed into Siegel-Narain theta-functions as described in section \ref{sec:mswcft}. One may thus ask the question what it implies for the functions $\hat f^{(r)}_{\mu,J} (\tau)$ for general $r$. In order to extract this information we want to compare the coefficients in the theta-decomposition on both sides of (\ref{eq:anomalygeneral}). For this we need a generalization of the identity (\ref{theta-identity1}). A computation shows that
\be
\theta^{(n)}_{\nu,J}(\tau,z)\,\theta^{(r-n)}_{\lambda,J}(\tau,z)=\sum_{\mu\,\in\,\Lambda^*/\Lambda} c^\mu_{\nu\lambda}(\tau)\, \theta^{(r)}_{\mu,J}(\tau,z),
\ee
where $c^\mu_{\nu\lambda}$ are Siegel-Narain theta-functions themselves given by
\be
c^\mu_{\nu\lambda}(\tau)=\delta_g(\mu)\sum_{\xi\,\in\,\Lambda+\mu+\frac{g}{r}(\nu-\lambda)}\bar q^{-\frac{rn(r-n)}{2g^2}\xi_+^2}q^{\frac{rn(r-n)}{2g^2}\xi_-^2}
\ee
with $g=\text{gcd}(n,r-n)$ and $\delta_g(\mu)$ yields one if $r\mu$ is divisible by $g$ and vanishes otherwise. With this input one finds
\be\label{eq:holanomalyfmu}
\partial_{\bar\tau} \hat{f}^{(r)}_{\mu,J} (\tau)  \sim \sum_{n=1}^{r-1}n(r-n)\sum_{\nu,\lambda\,\in\Lambda^*/\Lambda} \hat{f}^{(n)}_{\nu,J}(\tau) \hat{f}^{(r-n)}_{\lambda,J}(\tau) c^\mu_{\nu\lambda}(\tau),
\ee
which sheds some light into the question about the modular properties of generating functions at higher rank as follows.

The structure of eq.~(\ref{eq:holanomalyfmu}) indicates, that an appropriate description of the generating function $\hat{f}^{(r)}_{\mu,J}$ needs a generalization of the usual notion of mock modularity. This results from the fact, that on the right hand side of the anomaly equation (\ref{eq:holanomalyfmu}), mock modular forms appear, such that the shadow of $\hat{f}^{(r)}_{\mu,J}$ is a mock modular form itself. 
Therefore, it is also subject to a holomorphic anomaly equation. This would lead to the notion of mock modularity of higher depth \cite{Zwegers:higherdepth}, similar to the case of 
almost holomorphic modular forms of higher depth. These are functions like $\widehat{E}_2(\tau)$ and powers thereof, which can be written as a polynomial in $\tau_2^{-1}$ with coefficients 
being holomorphic functions.

A further motivation for this comes from the observation that the generating functions $\hat{f}^{(r)}_{\mu,J}$ could be obtained from an indefinite theta-function as in the case of two M5-branes. The lattice, however, that is summed over in these higher rank indefinite theta-functions will be of higher signature. In the case of $r$ M5-branes one would expect a signature $(r-1, (r-1)(r(\Lambda)-1))$ due to the $r-1$ relative D2-brane charges of the possible $r$ decay products of D4-D2-D0 bound-states \cite{Manschot:2009ia,Manschot:2010xp}. However, a complete discussion of the modular properties of such functions and their relation to mock modular forms of depth is beyond the scope of this work. We would like to come back to this question in future research.

\subsubsection{The contour description}
\label{sec:contour}
The elliptic genus of $r$ M5-branes wrapping $P$ is denoted by $Z^{(r)}_{P}(\tau,z)$, where we don't indicate any dependence of $Z^{(r)}_{P}$ on a K\"ahler class/ chamber $J\in\cC(P)$. The basic assumption is that the elliptic genus does not depend on such a choice. We simply think about $Z^{(r)}_{P}$ as being a \textit{meromorphic} Jacobi form, which has poles as a function of the elliptic variable $z$. We assume, that it is of bi-weight $(-\frac{3}{2},\frac{1}{2})$. In the following we want to exploit the implications of this statement.

It is known that a Jacobi form has an expansion into theta-functions with coefficients being modular forms. Since Zwegers \cite{Zwegers:2002}, we also know that a meromorphic Jacobi form with one elliptic variable has a similar expansion, where the coefficients are mock modular. Using our Siegel-Narain theta-function $\theta_{\mu,J}^{(r)}(\tau,z)$, eq.~(\ref{SNTheta}), we conjecture the following expansion
\be\label{decompositionZ}
Z^{(r)}_{P}(\tau,z)=\sum_{\mu\,\in\,\Lambda^*/\Lambda} f^{(r)}_{\mu,J}(\tau)\theta_{\mu,J}^{(r)}(\tau,z)+\text{Res},
\ee
with $J$ a point in the K\"ahler cone which is related to a point $z_J\in\Lambda_\IC$ where the decomposition is carried out. Note, that in eq.~(\ref{decompositionZ}) the term ``Res'' should be given as a finite sum over the residues of $Z^{(r)}_{P}(\tau,z)$ in the fundamental domain $z_J+e\tau+e$ with $e=[0,1]^{r(\Lambda)}$.

Let's see how the dependence on $J$ comes about. Doing a Fourier transform we can write
\be
f^{(r)}_{\mu,J}(\tau)=(-1)^{r\mu\cdot[P]}\bar{q}^{\frac{r}{2}\mu_+^2}q^{-\frac{r}{2}\mu_-^2}\int_{{\cal C}_J}Z^{(r)}_{P}(\tau,z)e^{-2\pi ir(\mu+\frac{[P]}{2})\cdot z}dz,
\ee
where ${\cal C}_J$ is a contour which has to be specified since $Z^{(r)}_{P}$ is meromorphic. Due to the periodicity in the elliptic variable ${\cal C}_J$ can be given as $z_J+e$ for some point $z_J$. Now, suppose we have a parallelogram ${\cal P}=z_J+ez_{J'}+e$ and that there is a single pole of $Z^{(r)}_{P}$ inside $\cal P$, say at $z=z_0$. Then, we obtain by integrating over the boundary of $\cal P$
\be\label{eq:jumpingresidues}
f^{(r)}_{\mu,J}(\tau)-f^{(r)}_{\mu,J'}(\tau)=2\pi i\,\alpha_\mu(\tau)\,\underset{z=z_0}{\text{Res}}\left(Z^{(r)}_{P}(\tau,z)e^{-2\pi ir(\mu+\frac{[P]}{2})\cdot z}\right),
\ee
where we abbreviate
\be
\alpha_\mu(\tau)=(-1)^{r\mu\cdot[P]}\bar{q}^{\frac{r}{2}\mu_+^2}q^{-\frac{r}{2}\mu_-^2}.
\ee
That is, the coefficients of the Laurent expansion of the elliptic genus encode the jumping of the BPS numbers across walls of marginal stability and the walls are in one-to-one correspondence with the positions of the poles of $Z^{(r)}_{P}$. An analogous dependence on a contour of integration for wall-crossing of ${\cal N}=4$ dyons was introduced in refs.~\cite{Sen:2007vb,Cheng:2007ch}.

Moreover, the shadow of $f^{(r)}_{\mu,J}$ should be determined in terms of the residues of $Z^{(r)}_{P}$, since a generalizations of the ideas of \cite{Zwegers:2002} should show, that it is contained in the factor ``Res'' of eq.~(\ref{decompositionZ}). Thus, combining this result with the interpretation of eq.~(\ref{eq:jumpingresidues}) one expects, that the shadow not only renders $f^{(r)}_{\mu,J}$ modular, but also encodes the decay of bound-states and hence knows about the jumping of BPS invariants across walls of marginal stability.

It is tempting to speculate even further. When comparing our results to the case of dyon state counting in ${\cal N}=4$ theories \cite{Dabholkar:2007zz,Dabholkar:2010} 
one might suspect that there is an analog of the Igusa cusp form $\phi_{10}$ in our setup. In 
the ${\cal N}=4$ dyon case there are meromorphic Jacobi forms, often denoted $\psi_m$, which are summed up 
to give $\phi_{10}$. In analogy, it may be useful to introduce another parameter $\rho\in{\cal H}$ 
and to study the object
\be
\phi_{P}^{-1}(\tau,\rho,z)=\sum_{r\geq1}Z^{(r)}_{P}(\tau,z)e^{2\pi ir\rho}.
\ee

\section{Conclusions}

In this paper we investigated background dependence of theories that originate from  
$r$ M5-branes wrapping a smooth (semi-)rigid divisor $P$ in a Calabi-Yau three-fold background.  
Such divisors $P$ have $b_2^+=1$ and (semi-)positive anti-canonical class. 
In this case the wrapped M5-brane can be studied locally in the Calabi-Yau manifold
using an effective description of the M5-brane theory on $P\times T^2$ 
by a twisted $U(r)$ ${\cal N}=4$ Super-Yang-Mills theory on $P$. 

The main object of interest was the partition function $Z_{P}^{(r)}$ of 
the twisted gauge theory and its modular and holomorphic properties. 
This partition function can be related to the modified elliptic genus 
of the ${\cal N}=(0,4)$ sigma model description of 
the M5-brane. Using the spectral flow symmetry one 
establishes for all $r$ a decomposition of the partition function 
into vector-valued modular forms $\hat f^{(r)}_{\mu, J}(\tau)$ 
w.r.t.~the $S$-duality group of ${\cal N}=4$ Super-Yang-Mills 
and Siegel-Narain theta-functions $\theta^{(r)}_{\mu,J}(\tau,z)$. 

Our main result is a rigorous proof of a holomorphic anomaly equation 
of the partition function valid for rank two on all $P$ described above. 
The proof in section \ref{sec:holanomalyrantwo} relies on the large 
volume wall-crossing formula of G\"ottsche~\cite{Gottsche:1999} for 
invariants associated to sheaves on $P$, which are 
related to integer BPS invariants. By summing the change of the invariants  across 
all intermediate walls one can express the difference of the 
generating function of the invariants $f^{(r)}_{\mu, J}(\tau)$  in 
two arbitrary chambers $J$ and $J'$ in the K\"ahler cone in terms of 
an indefinite theta-function $\Theta^{J,J'}_{\Lambda,\mu}(\tau,z)$~\cite{Gottsche:1998}. 
This theta-function is regularized by cutting out the negative 
directions of the quadratic form on the homology lattice, 
a procedure which renders the result in general not modular. 
The spoiled $S$-duality invariance can be regained following 
the work of Zwegers by smoothing out the cutting procedure with 
the non-holomorphic error function. The non-holomorphicity introduced 
by this procedure completes the mock modular forms $f^{(r)}_{\mu, J}(\tau)$ 
to non-holomorphic modular forms $\hat f^{(r)}_{\mu, J}(\tau)$. 
The non-holomorphicity of the Siegel-Narain theta-functions on the other 
is trivial since it is annihilated by the  non-holomorphic 
heat operator. This allows to write a concise holomorphic 
anomaly equation for the partition function (\ref{eq:provenholan}).

We check this holomorphic anomaly equation and its implications 
for the counting of invariants of sheaves on $\mathbb{P}^2$, 
$\mathbb{F}_0$, $\mathbb{F}_1$ and ${\cal B}_8$ in section~\ref{sec:surfaceswithb+=1}.
The anomaly equation (\ref{eq:provenholan}) is in particular 
compatible with the form of a holomorphic anomaly that has been 
conjectured in the context of E-strings on $\frac{1}{2}$K3 for all $r$ 
and checked for certain classes using the duality to the genus zero topological 
string partition function~\cite{Minahan:1998vr}. Since the 
non-holomorphicity of the $\hat f^{(r)}_{\mu, J}(\tau)$ for $r>1$ is 
related in an intriguing way to mock modularity and wall-crossing, 
we analyzed the decomposition for arbitrary rank and give a general 
form of the conjectured general anomaly equation at the level of the 
$\hat f^{(r)}_{\mu, J}(\tau)$ in equation (\ref{eq:holanomalyfmu}), which 
indicates a theory of mock modular forms of higher depth \cite{Zwegers:higherdepth}.
The holomorphic limit of the $\hat f^{(r)}_{\mu, J}(\tau)$ 
yield generating functions for invariants associated to sheaves 
of rank $r$. However, it is in general difficult to provide
boundary conditions, which fix the holomorphic ambiguity.

The wall-crossing of G\"ottsche, which induces in the steps 
described above the non-holomorphicity of the $\hat f^{(2)}_{\mu, J}(\tau)$, 
can be rederived using the Kontsevich-Soibelman wall-crossing formula, 
as we did in section \ref{sec:KSGwallcrossing}. As the wall-crossing formula takes a primitive form at rank two, one can rewrite the generating function of BPS differences in terms of an indefinite theta-function. 
The Kontsevich-Soibelman formula can be used for arbitrary rank to determine 
the counting functions $f^{(r)}_{\mu, J}(\tau)$ for all sectors 
$\mu$ in all chambers, if it is known in one chamber for 
all $\mu$, e.g.~by a vanishing lemma or use of the 
blow-up formula. This was studied for rank 3 by~\cite{Manschot:2010nc}, 
where it was also shown that the rank three wall-crossing formula 
is primitive. In general if the wall-crossing formula is primitive,
the sum over walls induce lattice sums of 
signature $(r-1)(b_2^+,b_2^-)$ with similar regularization 
requirements as for the rank two case. It is an interesting 
question if the program of Zwegers to build modular objects 
can be extended to the higher rank situation and leads upon 
non-holomorphic modular completion to the conjectured form of 
holomorphic anomaly equation and a precise notion of the
mock modular forms of higher depth. 

The problem of providing boundary conditions at least 
in one chamber for the del Pezzo surfaces (except for 
the Hirzebruch surface $\mathbbm{F}_0$) can in principle be solved by 
using the blow formula in both directions in connection 
with the wall-crossing formula before and after the 
blow-up. However, the blow-up formula in the literature 
apply only if $r$ and $c_1\cdot J$ have no common 
divisor. This restriction forbids in general to 
provide boundary conditions for all sectors.

The higher genus information discussed in equation 
(\ref{eq:highergenus}) gives finer information 
about the cohomology of moduli spaces of sheaves 
than its Euler number. Namely, an elliptic genus 
obtained by tracing over the right $j^3_R$ quantum 
numbers of the Lefshetz decomposition in the cohomology 
of the moduli space. On rigid surfaces it can be further 
refined to include the general $\Omega$ background parameters 
of Nekrasov~\cite{Nekrasov:2002qd}, which capture the individual $(j^2_L,j^3_R)$ 
quantum numbers~\cite{Iqbal:2007ii}. For rank two such refined 
partition functions have been considered in \cite{Goettsche:2007}
and it should be possible to extend the consideration 
above to the refined BPS numbers. Furthermore, the relation between D6-D2-D0 brane systems as counted by topological string theory and D4-D2-D0 brane systems associated to black hole state counting is the hallmark of the OSV conjecture \cite{Ooguri:2004zv}, which has been intensively studied. Wall-crossing issues in combination with this conjecture have been studied in ref.~\cite{Denef:2007vg}  and more recently from an M-theory perspective for example in ref.~\cite{Aganagic:2009kf}. It would be interesting to examine the implications of the anomaly equation in these contexts.

A conceptually very interesting but at this point more speculative 
approach is to consider the elliptic genus as a $J$ independent 
meromorphic Jacobi form, as we did in section~\ref{sec:contour}. As shown by Zwegers such meromorphic Jacobi 
forms have an expansion in theta-functions whose coefficients 
are mock modular forms, just as holomorphic Jacobi forms have 
an expansion in theta-functions with holomorphic modular forms as
coefficients. This formalism relates the changes in the BPS numbers 
across walls of marginal stability to the different choices of the contour in the 
definition of $f^{(r)}_{\mu, J}(\tau)$ as a Fourier integral of 
$Z_{P}^{(r)}$, i.e.~to the poles in $Z_{P}^{(r)}$, like in the ${\cal N}=4$ case \cite{David:2006ru, Cheng:2007ch}.

\acknowledgments
We would like to thank Frederik Denef, Lothar G\"ottsche, Jan Manschot, Sameer Murthy, Cumrun Vafa, Stefan Vandoren, Bernard De Wit, Xi Yin and Don Zagier for valuable discussions and comments. We would also like to thank the organizers of the workshop ``Interfaces and wall-crossing'', 2009 at the Arnold Sommerfeld Center in Munich and the organizers of the workshop on ``Automorphic forms, Kac-Moody algebras and strings'' at the Max Planck Institute of Mathematics in Bonn for stimulating research on these topics. M.A.~and M.H.~are grateful to Peter Mayr for proposing the connection of E-strings and the M5-brane partition function and for many discussions. M.A.~and M.H.~would furthermore like to thank Jean Dominique L\"ange for joint work on related projects.  M.A.~is supported by the Hausdorff Center for Mathematics and DFG fellowship
AL 1407/1-1. The work of M.H.~is supported by the program ``Origin and Structure of the Universe''. The work of M.R.~and T.W.~is supported by the program ``Bonn-Cologne Graduate School of Physics and Astronomy".

\clearpage
\appendix

\section{Divisors in Calabi-Yau spaces}\label{geometry}

Let us recapitulate here some facts of the geometry of smooth divisors $P$  
in a Calabi-Yau three-fold $X$.

\subsection{General facts about rigid divisors} 
We start with some facts about complex 
surfaces. The Riemann Roch formula relates the signature $\sigma$ 
and arithmetic genus $\chi_0$ to Chern class integrals
\begin{equation}
\sigma=\sum_i (b^+_{2i}-b^-_{2i})= \frac{1}{3}\int_P( c_1^2-2c_2), \qquad \chi_0=\sum_i (-1)^i
h_{i,0}=\frac{1}{12}\int_P(c_1^2+c_2).
\label{signaturegenus}
\end{equation}

Regarding the embedding one has the distinction whether 
$P$ is very ample or not, i.e.~if the line bundle ${\cal L}_P$ 
is generated by its global sections or not. In the  former case $P$ has 
$h^0(X,{\cal L}_P)-1$ deformations and there exists an 
embedding $j: X\rightarrow \mathbb{P}^n_P$ so that  
${\cal L}_P=j^*({\cal O}(1))$, i.e.~$P$ can be described 
by some polynomial. This situation has been considered 
in~\cite{Maldacena:1997de}, where the deformations and 
$b^+,b^-$ have been given. Generically one has  $h^{2,0}(P)
=\frac{1}{2}(b_2^+-1)$, which is positive in the very ample 
case. 

In this work we consider mainly rigid smooth divisors. 
In this case one has no deformations and locally the 
Calabi-Yau manifold can be written as the total space of the canonical line 
bundle ${\cal O}(K_P)\rightarrow P$ and the latter can be globalized 
to a elliptic fibration over $P$, see section~\ref{toricdata}, for 
$P=\mathbb{F}_n$. In this case $\Lambda_P=\Lambda$, compare sec.~\ref{sec:mswcft}. 

As $X$ is a Calabi-Yau manifold and to allow no section, $P$ 
has to have a positive $D^2>0$ anti-canonical divisor class $D=-K_P$, 
which is also required to be nef, i.e.~$D.C\ge 0$ for any 
irreducible curve $C$. This defines a weak del Pezzo surface. If $D.C> 0$, 
then $D$ is ample and $P$ is a del Pezzo 
surface~\cite{delPezzo}. Del Pezzo surfaces are either $\mathcal{B}_n$, 
which are blow-ups of $\mathbb{P}^2$ in $n\leq8$ points 
or $\mathbb{P}^1\times \mathbb{P}^1$. We can also allow  
the Hirzebruch surface $\mathbb{F}_2$ which is weak del 
Pezzo. 

As $h_{1,0}=h_{2,0}=0$ one has $\chi_0(\mathcal{B}_n)=1$ for 
all surfaces under consideration. As the Euler number $\chi(\mathcal{B}_n)=
3+n$ one has by (\ref{signaturegenus}) that $\int_P c_1^2=9-n$, 
which implies that $n=9$ is the critical case for positive 
anti-canonical class, and  $(b_2^+,b_2^-)=(1,n)$. The case 
$n=9$ is called $\frac{1}{2}$K3. We include this semi-rigid 
situation.

In more detail the homology of $\mathcal{B}_n$ is generated by the 
hyperplane class $h$ of $\mathbb{P}^2$ and the exceptional 
divisors of the blow-ups $e_i$, with the non-vanishing 
intersections $h^2=1=-e_i^2$. The anti-canonical class is 
given by $-K_{\mathcal{B}_n}=3 h-\sum_{i=1}^n e_i$. 
Defining the lattice generated by this element in 
$H_2(P,\mathbb{Z})$ as $\mathbb{Z}_{K_{\mathcal{B}_n}}$ and 
$E^*_n= (\mathbb{Z}_{K_{\mathcal{B}_n}})^\perp$ one sees that 
$E^*_1$ is trivial and $E_n^*$ are the lattices of the Lie 
algebras $(A_1,A_1\times A_2, A_4,D_5, E_6,E_7,E_8)$ for 
$n=2,\ldots,8$. The corresponding basis in terms of $(h,e_i)$  
is worked out in~\cite{delPezzo}. The homology lattice for 
$B_9$ is $\Gamma^{1,1}\oplus E_8$, where $\Gamma^{1,1}$ is the 
hyperbolic lattice with standard metric.

In order to study topological string theory 
in Calabi-Yau backgrounds realized in simple toric ambient spaces, 
one has to consider situations in which $\Lambda 
\subset \Lambda_P$, which is the case for the 
$\frac{1}{2}$K3 realized in the toric ambient space
discussed in the next section.

\subsection{Toric data of CY containing Hirzebruch surfaces $\mathbbm{F}_n$}\label{toricdata}
Let $X$ be an elliptic fibration over $\mathbb{F}_n$ for $n=0,1,2$ given by a generic section of the anti-canonical bundle of the ambient spaces specified by the following vertices
\begin{equation*}
\begin{split}
D_0=(0,0,0,0), &\quad D_1=(0,0,0,1), \quad D_2=(0,0,1,0), \quad D_3=(0,0,-2,-3)\\
D_4=(0,-1,-2,-3),& \quad D_5=(0,1,-2,-3), \quad D_6=(1,0,-2,-3),\quad D_7=(-1,-n,-2,-3).
\end{split}
\end{equation*}
One finds large volume phases with the following Mori-vectors
\begin{center}
\begin{tabular}{c|c|ccccccc|c}
 & $D_0$ & $D_1$ & $D_2$ & $D_3$& $D_4$& $D_5$& $D_6$& $D_7$& \\
\hline
$l^1=$ & $-6$ & 3 & 2 & 1 & 0 & 0 &0 &0 & $C^1$ \\
$l^2=$ & 0 & 0 & 0 & $-2$ & 1 & 1 &0 &0 & $C^2$ \\
$l^3=$ & 0 & 0 & 0 & $n-2$ & $-n$ & 0 &1 &1 & $C^3$. \\
\end{tabular}
\end{center}
We choose a basis $\{C^A,A=1,2,3\}$ of $H_2(X,\IZ)$. Let $K_A$ be a Poincar\'e dual basis of the Chow group of linearly independent divisors of $X$, i.e.~$\int_{C^A}K_B=\delta^A_B$. The divisors $D_i=l_i^AK_A$ have intersections with the cycles $C^A$ given by $D_i.C^A=l_i^A$. We have the following non-vanishing intersections of the divisors given by
\be
K_1.K_2.K_3=1,\quad K_1.K_2^2=n,\quad K_1^2.K_2=n+2,\quad K_1^2.K_3=2,\quad K_1^3=8.
\ee
The divisor giving the Hirzebruch surface inside the Calabi-Yau manifold corresponds to
\be
[\mathbb{F}_n]=D_3=K_1-2K_2-(2-n)K_3.
\ee
Thus, the metric on $H^2({\mathbb F}_n,\IZ)$ coming from the intersections in the Calabi-Yau manifold is
\be
(K_A.K_B.[{\mathbb F}_n])=\begin{pmatrix} 0\, & 0\, & 0\, \\ 0\, & n\, & 1\, \\ 0\, & 1\, & 0\, \end{pmatrix}.
\ee
Projecting out the direction corresponding to the elliptic fiber we reduce the problem to the Hirzebruch surface itself. We denote by $F=K_3$ and $B=K_2-nK_3$ the class of the fiber and base, respectively. Thus, the canonical class reduces to $[{\mathbb F}_n]=-(2+n)F-2B$. The intersection numbers are given as follows
\be
\begin{pmatrix}F.F & F.B \\ B.F & B.B\end{pmatrix}=\begin{pmatrix}0 & 1 \\ 1 & -n\end{pmatrix}.
\ee
Hence, the K\"ahler cone is spanned by the two vectors $F$ and $2B+nF$, i.e.
\be
\cC(\mathbb{F}_n)=\{J\in H^2(\mathbb{F}_n,\IR)\,|\, J=t_1 F+ t_2 (2B+nF),\, t_1,t_2>0 \}.
\ee
For $n=1$ the geometry admits also an embedding of a K3 and a ${\cal B}_9$ surface.

\section{D-branes and sheaves}\label{Dbound}
In order to clarify our notation we collect some facts about D-brane charges and the stability conditions for a bound-state system of D4-D2-D0 branes wrapped around a divisor $i:P\hookrightarrow X$ inside a Calabi--Yau three-fold $X$. See e.g. \cite{Aspinwall:2004jr} for a review.

\subsection{Charges of D-branes and sheaves}\label{chargev}
The D4-D2-D0 brane-system is specified by a (coherent) sheaf $\mathcal{E}$ on $P$. The image of the K-theory charge of the sheaf $\mathcal{E}$ in $H^\text{even}(X,\mathbb{Q})$ is given by the Mukai vector \cite{Green:1996dd, Minasian:1997mm, Witten:1998cd}
\begin{equation}
\Gamma={\rm ch}(i_*\mathcal{E}) \sqrt{{\rm Td}(X)},
\end{equation}
where $i_*\mathcal{E}$ denotes the extension-sheaf to $X$. Using the Grothendieck-Riemann-Roch-theorem
\begin{equation}
i_*({\rm ch}(\mathcal{E})\,{\rm Td}(P))={\rm ch}(i_*E)\,{\rm Td}(X),
\end{equation}
and the expressions
\begin{align}
{\rm Td}(Y)^a&=1+\frac{a}{2}c_1(Y)+\frac{(3a^2-a) c_1(Y)^2+2a\,c_2(Y)}{24}\\
{\rm ch}(Y)&=\sum_{i=0}^3 {\rm ch}_i(Y) ={\rm rk}(Y) +c_1(Y)+\frac{1}{2}c_1(Y)^2-c_2(Y)\\
{\rm ch}(Y^*)&=\sum_{i=0}^3 {\rm ch}_i(Y^*) ={\rm rk}(Y) -c_1(Y)+\frac{1}{2}c_1(Y)^2-c_2(Y),
\end{align}
where $Y^*$ denotes the dual sheaf, one obtains \cite{Diaconescu:1999vp}:
\begin{equation}
\begin{split}
\Gamma&=r [P]+r\,i_*\left(\frac{c_1(\mathcal{E})}{r}+\frac{c_1(P)}{2}\right)\\
&\quad+r i_*\left(\frac{c_1(P)^2+c_2(P)}{12}+\frac{\frac{1}{2}(c_1(P)c_1(\mathcal{E})+c_1(\cE)^2)-c_2(\mathcal{E})}{r} \right)-\frac{c_2(X)\cdot [P]}{24},
\end{split}
\end{equation}
where $r$ is the rank of the sheaf $\mathcal{E}$ and one has to note, that $c_1(X)=0$ as $X$ is Calabi-Yau. Using the adjunction formula we arrive at
\begin{equation}
\Gamma=(Q_6\,,\,Q_4\,,\,Q_2\,,\,Q_0)=r \left(0\,,\,[P]\,,\,i_* F\,,\left[\,\frac{\chi(P)}{24}+\int_P\frac{1}{2}F^2-\Delta \right]\right).
\end{equation}
Here we introduced
\begin{align}
F&=\frac{c_1(\mathcal{E})}{r}+\frac{c_1(P)}{2},\\
\Delta&=\frac{1}{2r^2}\left(2r\, c_2(\mathcal{E})-(r-1)\,c_1(\mathcal{E})^2\right).
\end{align}
The quantity $\Delta$ is called the discriminant.

\subsection{Stability conditions and moduli space}\label{mustab}
\subsubsection*{\it $\Pi$-stability}
Given the K-theory charges the expression for the central charge from mirror symmetry is
\begin{equation}\label{eq:centralcharge}
\begin{split}
Z(\mathcal{E})=&-\int e^{-(B+iJ)}\, \Gamma(\mathcal{E})+({\rm instanton-corrections})\\
=&-\frac{r}{2}[P]\cdot t^2+t(i_*c_1(\mathcal{E})+\frac{r}{2}i_*c_1(P))-{\rm ch}_2(\mathcal{E})\\
&-\frac{1}{2}c_1(\mathcal{E})c_1(P)-\frac{r}{8}c_1(P)^2-\frac{r}{24}c_2(P)+\mathcal{O}(e^{-t}),
\end{split}
\end{equation}
where $J$ is the K\"ahler form of $X$ and $t=B+iJ$. We now denote the phase of $Z(\mathcal{E})$ by
\begin{equation}
\varphi(\mathcal{E})=\frac{1}{\pi}\,{\rm Arg}\, Z(\mathcal{E})=\frac{1}{\pi}\, {\rm Im} \log{Z}(\mathcal{E}).
\end{equation}
A sheaf $\mathcal{E}$ is called $\Pi$-(semi)-stable \cite{Douglas:2000gi, Douglas:2000ah} iff for every (well-behaved) subsheaf $\mathcal{F}$:
\begin{equation}
\varphi(\mathcal{F})\leq \varphi(\mathcal{E}),
\end{equation}
where the strict inequality amounts to stability. If the inequality is strictly fulfilled (a stable sheaf) a decay is impossible by charge and energy conservation. Note, that the $\Pi$-stability condition involves an infinite tower of quantum corrections at an arbitrary point in moduli space. 

\subsubsection*{\it $\mu$-stability} 
In a large volume phase $(t\rightarrow \infty)$ of the Calabi-Yau the instanton-corrections are suppressed by $\mathcal{O}(e^{-t})$ and the classical expressions become exact. In this limit we are left with \cite{Diaconescu:2007bf}:
\begin{equation}
\varphi(\mathcal{E})=\frac{1}{\pi}\,{\rm Im} \log \left( -\,\frac{r}{2}J^2\cdot [P] \right)+2\,\frac{J\cdot\hat{\mu}}{J^2\cdot[P]}+\mathcal{O}\left(\frac{1}{J^2}\right).
\end{equation}
$\Pi$-stability now amounts to the definition
\begin{equation}
(i^*J)\cdot \frac{c_1(\mathcal{F})}{{\rm rk}(\mathcal{F})} \leq (i^*J)\cdot \frac{c_1(\mathcal{E})}{{\rm rk}(\mathcal{E})}\quad \text{for any nice subsheaf}\quad \mathcal{F}\subseteq \mathcal{E},
\end{equation}
where $i^*J$ denotes the pullback of the K\"ahler form of the Calabi-Yau to $P$ and all expressions are understood on $P$. The quantity appearing in the above definition is called slope and denoted by $\mu(\mathcal{E})$, i.e.
\begin{equation}
\mu(\mathcal{E}):=(i^* J)\cdot \frac{c_1(\mathcal{E})}{{\rm rk}(\mathcal{E})}.
\end{equation}
The above condition is called $\mu$-(semi-)stability and the classical notion of the stringy $\Pi$-stability. Note also, that $\mu$-stability is not sensitive to how the lower dimensional charges are distributed among decay products. This is in contrast to $\Pi$-stability, where quantum corrections change this insensitivity.

\subsubsection*{\it Dimension of moduli space}
On general grounds the moduli space of a D-brane modelled by a sheaf $\mathcal{E}$ is given by ${\rm Ext}^1(\mathcal{E},\mathcal{E})$. The elements of this group count the number of marginal open string operators in the spectrum of the BCFT describing the B-brane. We assume, that $P$ is a rational surface and further that the sheaf $\mathcal{E}$ is $\mu$-stable and that $(i^*J)\cdot [K_P]\leq 0$. Under these assumptions the moduli space is smooth and the following formula for its dimension holds \cite{Maruyama:1977}
\begin{equation}
{\rm dim}\, {\rm Ext}^1(\mathcal{E},\mathcal{E})=1+r^2(2\Delta-1).
\end{equation}
A consequence is that for a slope-stable sheaf one has
\begin{equation}
\Delta\geq 0,
\end{equation}
which is a condition on the stable bundle's Chern classes.

\section{(Mock) modularity}\label{sec:mock}
In this appendix we want to review the definition and some basic properties of mock modular forms and give the definitions of modular forms appearing in the main body text.  For more details on mock modular forms the reader is referred to the mathematics' literature \cite{Zwegers:2002,Zagier:2007,MR2555930}.

\subsection{Mock modular forms}
Following \cite{Zagier:2007}, we denote the space of mock modular forms of weight $k$ by $\mathbb{M}_k$ and the space of modular forms by $M_k$. Mock modular forms are holomorphic functions of $\tau$, which is an element of the upper half plane $\mathcal{H}$, but do not transform in a modular covariant way. However, to every mock modular form $h$ of weight $k$ there exists a shadow $g\in M_{2-k}$ such that the function $\hat h$, given by
\begin{equation}\label{eq:mockshadow}
\hat h (\tau) = h(\tau)+ g^*(\tau)
\end{equation}
transforms as of weight $k$. Denoting by $g^c(z) = \overline{g(-\bar z)}$, the completion $g^*(\tau)$ is defined by
\begin{equation}
g^*(\tau) = -(2i)^{k} \int_{-\bar \tau}^\infty (z+\tau)^{-k} g^c(z)\, dz.
\end{equation}
Thus, $\hat h$ is modular but has a non-holomorphic dependence. The corresponding space containing forms of type \eqref{eq:mockshadow} is denoted by $\wh{\mathbb{M}}_k$. Given $g$ as the expansion $g(\tau)=\sum_{n \geq0} b_n q^n$, the completion $g^*(\tau)$ can also be written as
\begin{equation} \label{beta}
 g^*(\tau) = \sum_{n\geq0} n^{k-1} \ov{b}_n\, \beta_k(4n\tau_2)\, q^{-n},
\end{equation}
with $\tau_2=\textrm{Im}\,(\tau)$ and $\beta_k$ defined by
\begin{equation}
\beta_k(t)=\int_t^{\infty} u^{-k} e^{-\pi u} du.
\end{equation}

Conversely, given $\hat h$, one determines the shadow $g$ by taking the derivative of $\hat{h}$ with respect to $\bar{\tau}$. One easily sees that
\begin{equation}
\frac{\partial \hat{h}}{\partial \overline{\tau}} =  \frac{\partial g^*}{\partial \overline{\tau}} = \tau_2^{-k} \overline{g(\tau)}.
\end{equation}
This viewpoint opens another characterization of $\wh{\mathbb{M}}_k$ as the set of real-analytic functions $F$ that fulfill a certain differential equation. To be precise, let us define the space $\mathfrak{M}_k$ as the space of real-analytic functions $F$ in the upper half-plane $\mathcal{H}$ transforming as a modular form under $\Gamma\subset {\rm SL}(2,\mathbb{Z})$, i.e.
$$F(\gamma \tau) = \rho(\gamma)(c\tau+d)^k F(\tau), $$ where $\rho(\gamma)$ denotes some character of $\Gamma$ and we demand exponential growth at the cusps. Hence, the space of completed mock modular forms $\wh{\mathbb{M}}_k$ can now be characterized by 
\begin{equation}
\wh{\mathbb{M}}_k = \left\{F\in \mathfrak{M}_k \,\big|\, \frac{\partial}{\partial \tau}\left(\tau_2^k \frac{\partial F}{\partial \bar \tau}\right)=0\right\}.
\end{equation}
This definition induces the following maps\footnote{~A function $f\in \mathfrak{M}_{k,l}$ transforms under modular transformations $\gamma\in\Gamma$ with bi-weight $(k,l)$ and character $\rho$, i.e.~$f(\gamma \tau) = \rho(\gamma)(c\tau+d)^k (c \bar \tau +d)^l f(\tau)$.}
\begin{equation}
\mathfrak{M}_k=\mathfrak{M}_{k,0} \stackrel{\tau_2^k \partial_{\bar\tau}}{\longrightarrow}\mathfrak{M}_{0,2-k}\stackrel{\tau_2^{2-k} \partial_\tau}{\longrightarrow} \mathfrak{M}_{k,0}=\mathfrak{M}_k,
\end{equation}
so that the composition can be converted to the Laplace operator in weight $k$. Hence, mock modular forms in $\wh{\mathbb{M}}_k$ have the special eigenvalue $\frac{k}{2}\left(1-\frac{k}{2}\right)$ and are sometimes also called harmonic weak Maass forms.

Zwegers showed in \cite{Zwegers:2002} that mock modular forms can be realized in three different ways, namely either as Appell-Lerch sums, indefinite theta-series or as Fourier coefficients of meromorphic Jacobi forms. Further, there is a notion of mixed mock modular forms, which are functions that transform in the tensor space of mock modular forms and modular forms. However, we will call them simply mock modular forms as well.

In the following a simple example of a mock modular form is presented.

\subsubsection*{\it Example: $E_2$ as a mock modular form}
The modular completion of the holomorphic Eisenstein series $E_2$ (see below for a definition) has the form
$$ \widehat{E_2}(\tau) = E_2(\tau) -\frac{3}{\pi \tau_2} \, .$$
From $\partial_{\tb} \wh{E}_2 = \tau_2^{-2} \frac{3i}{2\pi}$ we get $\ov{g} = \frac{3i}{2\pi}$, a constant shadow. Doing the integral indeed yields
\begin{equation}
g^*(\tau) = -\left(2 i\right)^{2} \int_{-\tb}^{\infty} (z+ \tau)^{-2} \frac{3i}{2 \pi} dz = -\frac{6 i}{\pi} \left[\frac{-1}{z+\tau}\right]_{-\tb}^{\infty} = - \frac{3}{\pi \tau_2}.
\end{equation}

\subsection{Modular forms}
Let us collect the definitions of various modular forms appearing in the main body text. We denote the following standard theta-functions by
\begin{equation}
\begin{split}
\vartheta_1(\tau,\nu)&=\sum_{n\in\mathbb{Z}+\frac{1}{2}}(-1)^n q^{\frac{1}{2}n^2}e^{2\pi in\nu},\\
\vartheta_2(\tau,\nu)&=\sum_{n\in\mathbb{Z}+\frac{1}{2}} q^{\frac{1}{2}n^2}e^{2\pi in\nu},\\
\vartheta_3(\tau,\nu)&=\sum_{n\in\mathbb{Z}} q^{\frac{1}{2}n^2}e^{2\pi in\nu},\\
\vartheta_4(\tau,\nu)&=\sum_{n\in\mathbb{Z}}(-1)^n q^{\frac{1}{2}n^2}e^{2\pi in\nu}.
\end{split}
\end{equation}
In the case that $\nu=0$ we simply denote $\vartheta_i(\tau)=\vartheta_i(\tau,0)$ (notice that $\vartheta_1(\tau)=0$). Under modular transformations the theta functions $\vartheta_i(\tau)$ behave as vector-valued modular forms of weight $\frac{1}{2}$. They transform as
\begin{eqnarray}
\vartheta_2(-1/\tau)=\sqrt{\frac{\tau}{i}}\vartheta_4(\tau),&\quad& \vartheta_2(\tau+1)=e^{\frac{i\pi}{4}}\vartheta_2(\tau), \\
\vartheta_3(-1/\tau)=\sqrt{\frac{\tau}{i}}\vartheta_3(\tau),&\quad& \vartheta_3(\tau+1)=\vartheta_4(\tau), \\
\vartheta_4(-1/\tau)=\sqrt{\frac{\tau}{i}}\vartheta_2(\tau),&\quad& \vartheta_4(\tau+1)=\vartheta_3(\tau).
\end{eqnarray}
Further, the eta-function is defined by
\begin{equation}
\eta(\tau)=q^{\frac{1}{24}}\prod_{n=1}^\infty(1-q^n),
\end{equation}
and transforms according to
\begin{equation}\label{etatrafo}
\eta(\tau+1)=e^{\frac{i\pi}{12}}\eta(\tau),\qquad \eta\left(-\frac{1}{\tau}\right)=\sqrt{\frac{\tau}{i}}\,\eta(\tau).
\end{equation}
The Eisenstein series are defined by
\begin{equation}\label{eisensteinseries}
E_k(\tau)=1-\frac{2k}{B_k}\sum_{n=1}^\infty\frac{n^{k-1}q^n}{1-q^n},
\end{equation}
where $B_k$ denotes the $k$-th Bernoulli number. $E_k$ is a modular form of weight $k$ for $k>2$ and even.

\subsection{Modular properties of the elliptic genus}\label{app:modularity}
We denote by $Z^{(r)}_{P}(\tau,z)$ the elliptic genus of $r$ M5-branes wrapping $P$ as defined previously in sect.~\ref{sec:mswcft}. The elliptic genus should transform like a Jacobi form of bi-weight $(-\frac{3}{2},\frac{1}{2})$ and bi-index $(\frac{r}{2}(d_{AB}-\frac{J_AJ_B}{J^2}),\frac{r}{2}\frac{J_AJ_B}{J^2})$ under the full modular group. In particular, we impose
\be
\begin{split}\label{ellgentrafo}
Z^{(r)}_{P}(\tau+1,z)&=\varepsilon(T)\,Z^{(r)}_{P}(\tau,z),\\
Z^{(r)}_{P}(-\frac{1}{\tau},\frac{z_-}{\tau}+\frac{z_+}{\bar\tau})&=\varepsilon(S)\,
\tau^{-\frac{3}{2}}\bar\tau^{\frac{1}{2}}e^{\pi i r(\frac{z_-^2}{\tau}+\frac{z_+^2}{\bar\tau})}\,Z^{(r)}_{P}(\tau,z),
\end{split}
\ee
where $\varepsilon$ are certain phases \cite{Manschot:2008zb}.

\subsection*{\it Siegel-Narain theta-function and its properties}
Let us start by recalling the definition of the Siegel-Narain theta-function of eq.~(\ref{SNTheta})
\be
\theta_{\mu,J}^{(r)}(\tau,z)=\sum_{\xi\,\in\,\Lambda+\frac{[P]}{2}}(-)^{r(\xi+\mu)\cdot[P]}\bar{q}^{-\frac{r}{2}(\xi+\mu)_+^2}q^{\frac{r}{2}(\xi+\mu)_-^2}e^{2\pi ir(\xi+\mu)\cdot z},
\ee
where we define
\be
\xi_+^2=\frac{(\xi\cdot J)^2}{J\cdot J},\quad \xi_-^2=\xi^2-\xi_+^2.
\ee
Note, that $\xi_+^2<0$ if $J$ lies in the K\"ahler cone. 

If we denote by ${\cal D}_k=\p_{\bar\tau}+\frac{i}{4\pi k}\p^2_{z_+}$, the theta-function fulfills the heat equation
\be
{\cal D}_r\,\theta_{\mu,J}^{(r)}(\tau,z)=0.
\ee
Further, we denote by $\Lambda^*$ the dual lattice to $\Lambda$ w.r.t.~the metric $rd_{AB}$. For $\mu\in\Lambda^*/\Lambda$, we can deduce the following set of transformation rules
\be
\begin{split}\label{thetatrafo}
\theta_{\mu,J}^{(r)}(\tau+1,z)&=(-1)^{r(\mu+\frac{[P]}{2})^2}\theta_{\mu,J}^{(r)}(\tau,z),\\
\theta_{\mu,J}^{(r)}(-\frac{1}{\tau},\frac{z_+}{\bar\tau}+\frac{z_-}{\tau})&=\frac{(-1)^{r\frac{[P]^2}{2}}}{\sqrt{|\Lambda^*/\Lambda|}}(-i\tau)^{\frac{r(\Lambda)-1}{2}}(i\bar\tau)^{\frac{1}{2}}e^{\pi ir(\frac{z_-^2}{\tau}+\frac{z_+^2}{\bar\tau})}\,\sum_{\delta\,\in\,\Lambda^*/\Lambda}e^{-2\pi ir\mu\cdot\delta}\theta_{\delta,J}^{(r)}(\tau,z).
\end{split}
\ee

\subsection*{\it Rank one}
At rank one we have the universal answer
\be
f^{(1)}_{\mu,J}(\tau)=\frac{\vartheta_{\Lambda^\perp}(\tau)}{\eta(\tau)^\chi}.
\ee
The transformation rules are simply given by (\ref{etatrafo}) for the eta-function and  for $\vartheta_{\Lambda^\perp}$ we obtain (assuming $\Lambda^\perp$ even and self-dual)
\be
\begin{split}
\vartheta_{\Lambda^\perp}(\tau+1)&=\vartheta_{\Lambda^\perp}(\tau),\\
\vartheta_{\Lambda^\perp}(-\frac{1}{\tau})&=\left(\frac{\tau}{i}\right)^{\frac{r(\Lambda^\perp)}{2}}\vartheta_{\Lambda^\perp}(\tau).
\end{split}
\ee

\subsection*{\it Rank two}
Using Zwegers' theta-function with characteristics $\vartheta^{c,c'}_{a,b}(\tau)$ given in def.~2.1 of his thesis \cite{Zwegers:2002}, we can write
\be
\widehat{\Theta}^{c,c'}_{\Lambda,\mu}(\tau,x)=q^{-\frac{1}{2}\langle a,a\rangle}e^{-2\pi i\langle a,b\rangle}\vartheta^{c,c'}_{a+\mu,b}(\tau),
\ee
where $x=a\tau+b$, i.e.
\be
a=\frac{\text{Im}(x)}{\text{Im}(\tau)},\quad b=\frac{\text{Im}(\bar{x}\tau)}{\text{Im}(\tau)}.
\ee
Following Corollary 2.9 of Zwegers \cite{Zwegers:2002}, we can deduce the following set of transformations
\be
\begin{split}
\widehat{\Theta}^{c,c'}_{\Lambda,\mu}(\tau+1,x)&=(-1)^{\langle\mu,\mu\rangle}\widehat{\Theta}^{c,c'}_{\Lambda,\mu}(\tau,x), \\
\widehat{\Theta}^{c,c'}_{\Lambda,\mu}(-\frac{1}{\tau},\frac{x}{\tau})&=\frac{i(-i\tau)^{r(\Lambda)/2}}{\sqrt{|\Lambda^*/\Lambda|}}e^{\pi i\frac{\langle x,x\rangle}{\tau}}\sum_{\delta\,\in\,\Lambda^*/\Lambda}e^{-2\pi i\langle\delta,\mu\rangle}\,\widehat{\Theta}^{c,c'}_{\Lambda,\delta}(\tau,x).
\end{split}
\ee
This input enables us to write down the transformation rules for $\hat{f}^{(2)}_{\mu,J}$. They read
\be
\begin{split}
\hat{f}^{(2)}_{\mu,J}(\tau+1)&=(-1)^{\frac{\chi}{6}+2\mu^2}\hat{f}^{(2)}_{\mu,J}(\tau), \\
\hat{f}^{(2)}_{\mu,J}(-\frac{1}{\tau})&=-\frac{(-i\tau)^{-\frac{r(\Lambda)+2}{2}}}{\sqrt{|\Lambda^*/\Lambda|}}\sum_{\delta\,\in\,\Lambda^*/\Lambda}e^{4\pi i\delta\cdot\mu}\hat{f}^{(2)}_{\delta,J}(\tau).
\end{split}
\ee
This gives the conjectured transformation properties (\ref{ellgentrafo}).

\subsection*{\it The blow-up factor}
For completeness we elaborate on the transformation properties of the blow-up factor. We define
\be
\begin{split}
B_{r,k}(\tau)&=\eta(\tau)^{-r} \sum_{a_i\,\in\,\IZ+\frac{k}{r}}q^{\sum_{i\leq j\leq r-1}a_ia_j}.
\end{split}
\ee
We can deduce the following set of transformation rules
\be
\begin{split}
B_{r,k}(\tau+1)&=(-1)^{\frac{r}{12}+\frac{k^2(r-1)}{r}}B_{r,k}(\tau), \\[0.25cm]
B_{r,k}(-\frac{1}{\tau})&=\frac{1}{\sqrt{r}}\left(\frac{\tau}{i}\right)^{-\frac{1}{2}}\sum_{0\leq l\leq r-1}(-1)^{\frac{2kl(r-1)}{r}}B_{r,l}(\tau). \\
\end{split}
\ee

\section{Elliptic genera of K3 and $\frac{1}{2}$K3}\label{app:K3}
In the following we give some further examples of elliptic genera of multiple M5-branes wrapping the K3 and $\frac{1}{2}$K3 surfaces within the geometry of ref.~\cite{Klemm:1996hh}. The expressions for the elliptic genera can be read off from the instanton part of the prepotential of the geometry (see section \ref{n4ym}) and were given in ref.~\cite{Hecht:2008}, the $\frac{1}{2}$K3 expressions were known previously in refs.~\cite{Minahan:1997ct,Minahan:1998vr}. 

\subsection*{\it Elliptic genera of multiply wrapping the K3}
These are obtained by setting $q_2\rightarrow 0$ and can all be obtained from $Z^{(1)}$ by the Hecke transformation. 
\begin{eqnarray*}
Z^{(1)}&=&-\frac{2 E_4 E_6 }{\eta^{24}}\\
Z^{(2)}&=&-\frac{E_4 E_6 \left(17 E_4^3+7 E_6^2\right) }{96 \eta^{48}}\\
Z^{(3)}&=&-\frac{\left(9349 E_4^7 E_6+16630 E_4^4 E_6^3+1669 E_4 E_6^5\right) }{373248 \eta^{72}}\\
Z^{(4)}&=&-\frac{E_4 E_6 \left(11422873 E_4^9+46339341 E_4^6 E_6^2+21978651 E_4^3 E_6^4+880703 E_6^6\right)
}{2579890176 \eta^{96}}\\
Z^{(5)}&=&-\frac{E_4 E_6 \left(27411222535 E_4^{12}+198761115620 E_4^9 E_6^2+222886195242 E_4^6 E_6^4\right)}{30958682112000 \eta^{120}}\\
&&- \frac{E_4 E_6\left(45368414180 E_4^3 E_6^6+911966215 E_6^8\right) }{30958682112000 \eta^{120}}
\end{eqnarray*}

\subsection*{\it Elliptic genera of $\frac{1}{2}$K3, E-string bound-states}
These are obtained by setting $q_3\rightarrow 0$, the polynomials containing $E_2$ represent the part coming from bound-states. The polynomial appearance of $E_2$ at higher wrapping is an example of the appearance of mock modular forms of higher depth at higher wrapping.

\begin{eqnarray*}
Z^{(1)} &=&\frac{E_4 \sqrt{q}}{\eta^{12}}\\
Z^{(2)} &=&\frac{E_4 (E_2 E_4+2 E_6) q}{24 \eta^{24}}\\
Z^{(3)} &=& \frac{E_4 \left(54 E_2^2 E_4^2+109 E_4^3+216 E_2 E_4 E_6+197 E_6^2\right) q^{3/2}}{15552 \eta^{36}}\\
Z^{(4)} &=&\frac{E_4 \left(24 E_2^3 E_4^3+109 E_2 E_4^4+144 E_2^2 E_4^2 E_6+272 E_4^3 E_6+269
E_2 E_4 E_6^2+154 E_6^3\right) q^2}{62208 \eta^{48}}\\
Z^{(5)} &=& \frac{E_4  \left(18750 E_2^4 E_4^4+150000 E_2^3 E_4^3 E_6+1250 E_2^2 \left(109 E_4^5+341 E_4^2
   E_6^2\right)\right) q^{5/2}}{373248000 \eta^{60}}
   \\
   &+&\frac{E_4  \left(1000 E_2 \left(653 E_4^4 E_6+505 E_4 E_6^3\right)+116769 E_4^6+772460 E_4^3
   E_6^2+207505 E_6^4\right)q^{5/2}}{373248000 \eta^{60}}
  \end{eqnarray*}

\newpage
\bibliography{AHHKRW}

\providecommand{\href}[2]{#2}\begingroup\raggedright\begin{thebibliography}{10%
0}

\bibitem{Witten:1988xj}
E.~Witten, {\it {Topological Sigma Models}},  {\em Commun. Math. Phys.} {\bf
  118} (1988) 411.

\bibitem{Gopakumar:1998ii}
R.~Gopakumar and C.~Vafa, {\it {M-theory and topological strings. I}},
  \href{http://xxx.lanl.gov/abs/hep-th/9809187}{{\tt hep-th/9809187}}.

\bibitem{Gopakumar:1998jq}
R.~Gopakumar and C.~Vafa, {\it {M-theory and topological strings. II}},
  \href{http://xxx.lanl.gov/abs/hep-th/9812127}{{\tt hep-th/9812127}}.

\bibitem{Minahan:1998vr}
J.~A. Minahan, D.~Nemeschansky, C.~Vafa, and N.~P. Warner, {\it {E-strings and
  N = 4 topological Yang-Mills theories}},  {\em Nucl. Phys.} {\bf B527} (1998)
  581--623, [\href{http://xxx.lanl.gov/abs/hep-th/9802168}{{\tt
  hep-th/9802168}}].

\bibitem{Vafa:1994tf}
C.~Vafa and E.~Witten, {\it {A Strong coupling test of S duality}},  {\em Nucl.
  Phys.} {\bf B431} (1994) 3--77,
  [\href{http://xxx.lanl.gov/abs/hep-th/9408074}{{\tt hep-th/9408074}}].

\bibitem{Gaiotto:2006wm}
D.~Gaiotto, A.~Strominger, and X.~Yin, {\it {The M5-brane elliptic genus:
  Modularity and BPS states}},  {\em JHEP} {\bf 08} (2007) 070,
  [\href{http://xxx.lanl.gov/abs/hep-th/0607010}{{\tt hep-th/0607010}}].

\bibitem{Maldacena:1997de}
J.~M. Maldacena, A.~Strominger, and E.~Witten, {\it {Black hole entropy in
  M-theory}},  {\em JHEP} {\bf 12} (1997) 002,
  [\href{http://xxx.lanl.gov/abs/hep-th/9711053}{{\tt hep-th/9711053}}].

\bibitem{Bershadsky:1993ta}
M.~Bershadsky, S.~Cecotti, H.~Ooguri, and C.~Vafa, {\it {Holomorphic anomalies
  in topological field theories}},  {\em Nucl. Phys.} {\bf B405} (1993)
  279--304, [\href{http://xxx.lanl.gov/abs/hep-th/9302103}{{\tt
  hep-th/9302103}}].

\bibitem{Bershadsky:1993cx}
M.~Bershadsky, S.~Cecotti, H.~Ooguri, and C.~Vafa, {\it {Kodaira-Spencer theory
  of gravity and exact results for quantum string amplitudes}},  {\em Commun.
  Math. Phys.} {\bf 165} (1994) 311--428,
  [\href{http://xxx.lanl.gov/abs/hep-th/9309140}{{\tt hep-th/9309140}}].

\bibitem{Yamaguchi:2004bt}
S.~Yamaguchi and S.-T. Yau, {\it {Topological string partition functions as
  polynomials}},  {\em JHEP} {\bf 07} (2004) 047,
  [\href{http://xxx.lanl.gov/abs/hep-th/0406078}{{\tt hep-th/0406078}}].

\bibitem{Grimm:2007tm}
T.~W. Grimm, A.~Klemm, M.~Marino, and M.~Weiss, {\it {Direct integration of the
  topological string}},  {\em JHEP} {\bf 08} (2007) 058,
  [\href{http://xxx.lanl.gov/abs/hep-th/0702187}{{\tt hep-th/0702187}}].

\bibitem{Alim:2007qj}
M.~Alim and J.~D. Lange, {\it {Polynomial Structure of the (Open) Topological
  String Partition Function}},  {\em JHEP} {\bf 10} (2007) 045,
  [\href{http://xxx.lanl.gov/abs/0708.2886}{{\tt arXiv:0708.2886}}].

\bibitem{Huang:2006si}
M.-x. Huang and A.~Klemm, {\it {Holomorphic anomaly in gauge theories and
  matrix models}},  {\em JHEP} {\bf 09} (2007) 054,
  [\href{http://xxx.lanl.gov/abs/hep-th/0605195}{{\tt hep-th/0605195}}].

\bibitem{Huang:2006hq}
M.-x. Huang, A.~Klemm, and S.~Quackenbush, {\it {Topological String Theory on
  Compact Calabi-Yau: Modularity and Boundary Conditions}},  {\em Lect. Notes
  Phys.} {\bf 757} (2009) 45--102,
  [\href{http://xxx.lanl.gov/abs/hep-th/0612125}{{\tt hep-th/0612125}}].

\bibitem{Haghighat:2008gw}
B.~Haghighat, A.~Klemm, and M.~Rauch, {\it {Integrability of the holomorphic
  anomaly equations}},  {\em JHEP} {\bf 10} (2008) 097,
  [\href{http://xxx.lanl.gov/abs/0809.1674}{{\tt arXiv:0809.1674}}].

\bibitem{Alim:2008kp}
M.~Alim, J.~D. Lange, and P.~Mayr, {\it {Global Properties of Topological
  String Amplitudes and Orbifold Invariants}},  {\em JHEP} {\bf 03} (2010) 113,
  [\href{http://xxx.lanl.gov/abs/0809.4253}{{\tt arXiv:0809.4253}}].

\bibitem{Witten:1993ed}
E.~Witten, {\it {Quantum background independence in string theory}},
  \href{http://xxx.lanl.gov/abs/hep-th/9306122}{{\tt hep-th/9306122}}.

\bibitem{Aganagic:2006wq}
M.~Aganagic, V.~Bouchard, and A.~Klemm, {\it {Topological Strings and (Almost)
  Modular Forms}},  {\em Commun. Math. Phys.} {\bf 277} (2008) 771--819,
  [\href{http://xxx.lanl.gov/abs/hep-th/0607100}{{\tt hep-th/0607100}}].

\bibitem{Dijkgraaf:2002ac}
R.~Dijkgraaf, E.~P. Verlinde, and M.~Vonk, {\it {On the partition sum of the NS
  five-brane}},  \href{http://xxx.lanl.gov/abs/hep-th/0205281}{{\tt
  hep-th/0205281}}.

\bibitem{Verlinde:2004ck}
E.~P. Verlinde, {\it {Attractors and the holomorphic anomaly}},
  \href{http://xxx.lanl.gov/abs/hep-th/0412139}{{\tt hep-th/0412139}}.

\bibitem{Gunaydin:2006bz}
M.~Gunaydin, A.~Neitzke, and B.~Pioline, {\it {Topological wave functions and
  heat equations}},  {\em JHEP} {\bf 12} (2006) 070,
  [\href{http://xxx.lanl.gov/abs/hep-th/0607200}{{\tt hep-th/0607200}}].

\bibitem{Zagier:1975}
D.~Zagier, {\it {Nombres de classes et formes modulaires de poids $3/2$}},
  {\em C. R. Acad. Sci. Paris.} {\bf 21} (1975) A883--A886.

\bibitem{Minahan:1997ct}
J.~A. Minahan, D.~Nemeschansky, and N.~P. Warner, {\it {Partition functions for
  BPS states of the non-critical E(8) string}},  {\em Adv. Theor. Math. Phys.}
  {\bf 1} (1998) 167--183, [\href{http://xxx.lanl.gov/abs/hep-th/9707149}{{\tt
  hep-th/9707149}}].

\bibitem{Zwegers:2002}
S.~P. Zwegers, {\it {Mock Theta Functions}},  {\em Proefschrift Universiteit
  Utrecht} (2002).

\bibitem{Zagier:2007}
D.~Zagier, {\it {Ramanujan's Mock Theta Functions and their Applications
  d'apr\`es Zwegers and Bringmann-Ono}},  {\em S\'eminaire BOURBAKI} {\bf 986}
  (2007).

\bibitem{MR2555930}
K.~Ono, {\it Unearthing the visions of a master: harmonic maass forms and
  number theory},  {\em Current developments in mathematics} {\bf 2008} (2009)
  347--454.

\bibitem{Gottsche:1998}
L.~Gottsche and D.~Zagier, {\it {Jacobi forms and the structure of Donaldson
  invariants for 4-manifolds with $b_+=1$}},  {\em Sel. math.} {\bf New ser. 4}
  (1998) 69--115.

\bibitem{Moore:1997pc}
G.~W. Moore and E.~Witten, {\it {Integration over the u-plane in Donaldson
  theory}},  {\em Adv. Theor. Math. Phys.} {\bf 1} (1998) 298--387,
  [\href{http://xxx.lanl.gov/abs/hep-th/9709193}{{\tt hep-th/9709193}}].

\bibitem{Losev:1997tp}
A.~Losev, N.~Nekrasov, and S.~L. Shatashvili, {\it {Issues in topological gauge
  theory}},  {\em Nucl. Phys.} {\bf B534} (1998) 549--611,
  [\href{http://xxx.lanl.gov/abs/hep-th/9711108}{{\tt hep-th/9711108}}].

\bibitem{Cecotti:1992rm}
S.~Cecotti and C.~Vafa, {\it {On classification of N=2 supersymmetric
  theories}},  {\em Commun. Math. Phys.} {\bf 158} (1993) 569--644,
  [\href{http://xxx.lanl.gov/abs/hep-th/9211097}{{\tt hep-th/9211097}}].

\bibitem{Seiberg:1994rs}
N.~Seiberg and E.~Witten, {\it {Monopole Condensation, And Confinement In N=2
  Supersymmetric Yang-Mills Theory}},  {\em Nucl. Phys.} {\bf B426} (1994)
  19--52, [\href{http://xxx.lanl.gov/abs/hep-th/9407087}{{\tt
  hep-th/9407087}}].

\bibitem{Denef:2000nb}
F.~Denef, {\it {Supergravity flows and D-brane stability}},  {\em JHEP} {\bf
  08} (2000) 050, [\href{http://xxx.lanl.gov/abs/hep-th/0005049}{{\tt
  hep-th/0005049}}].

\bibitem{Denef:2007vg}
F.~Denef and G.~W. Moore, {\it {Split states, entropy enigmas, holes and
  halos}},  \href{http://xxx.lanl.gov/abs/hep-th/0702146}{{\tt
  hep-th/0702146}}.

\bibitem{KS:2008}
Kontsevich and Soibelman, {\it {Stability structures, motivic Donaldson-Thomas
  invariants and cluster transformations}},
  \href{http://xxx.lanl.gov/abs/0811.2435}{{\tt arXiv:0811.2435}}.

\bibitem{Gaiotto:2008cd}
D.~Gaiotto, G.~W. Moore, and A.~Neitzke, {\it {Four-dimensional wall-crossing
  via three-dimensional field theory}},  {\em Commun. Math. Phys.} {\bf 299}
  (2010) 163--224, [\href{http://xxx.lanl.gov/abs/0807.4723}{{\tt
  arXiv:0807.4723}}].

\bibitem{Cecotti:2009uf}
S.~Cecotti and C.~Vafa, {\it {BPS Wall Crossing and Topological Strings}},
  \href{http://xxx.lanl.gov/abs/0910.2615}{{\tt arXiv:0910.2615}}.

\bibitem{Gaiotto:2009hg}
D.~Gaiotto, G.~W. Moore, and A.~Neitzke, {\it {Wall-crossing, Hitchin Systems,
  and the WKB Approximation}},  \href{http://xxx.lanl.gov/abs/0907.3987}{{\tt
  arXiv:0907.3987}}.

\bibitem{Gaiotto:2010be}
D.~Gaiotto, G.~W. Moore, and A.~Neitzke, {\it {Framed BPS States}},
  \href{http://xxx.lanl.gov/abs/1006.0146}{{\tt arXiv:1006.0146}}.

\bibitem{Cecotti:2010fi}
S.~Cecotti, A.~Neitzke, and C.~Vafa, {\it {R-Twisting and 4d/2d
  Correspondences}},  \href{http://xxx.lanl.gov/abs/1006.3435}{{\tt
  arXiv:1006.3435}}.

\bibitem{Dabholkar:2007zz}
A.~Dabholkar, {\it {Cargese lectures on black holes, dyons, and modular
  forms}},  {\em Nucl. Phys. Proc. Suppl.} {\bf 171} (2007) 2--15.

\bibitem{Dabholkar:2010}
A.~Dabholkar, S.~Murthy, and D.~Zagier, ``Quantum black holes and mock modular
  forms.'' Talks at ASC workshop on Interfaces and Wall crossing, 2009 in
  Munich, workshop on Automorphic Forms, Kac-Moody Algebras and Strings, 2010
  in Bonn and at the conference on Topological String Theory, Modularity and
  Non-perturbative Physics, 2010 in Vienna.

\bibitem{Manschot:2009ia}
J.~Manschot, {\it {Stability and duality in N=2 supergravity}},
  \href{http://xxx.lanl.gov/abs/0906.1767}{{\tt arXiv:0906.1767}}.

\bibitem{Manschot:2010xp}
J.~Manschot, {\it {Wall-crossing of D4-branes using flow trees}},
  \href{http://xxx.lanl.gov/abs/1003.1570}{{\tt arXiv:1003.1570}}.

\bibitem{Manschot:2010nc}
J.~Manschot, {\it {The Betti numbers of the moduli space of stable sheaves of
  rank 3 on P2}},  \href{http://xxx.lanl.gov/abs/1009.1775}{{\tt
  arXiv:1009.1775}}.

\bibitem{Bringmann:2010sd}
K.~Bringmann and J.~Manschot, {\it {From sheaves on P2 to a generalization of
  the Rademacher expansion}},  \href{http://xxx.lanl.gov/abs/1006.0915}{{\tt
  arXiv:1006.0915}}.

\bibitem{Eguchi:2008gc}
T.~Eguchi and K.~Hikami, {\it {Superconformal Algebras and Mock Theta
  Functions}},  {\em J. Phys.} {\bf A42} (2009) 304010,
  [\href{http://xxx.lanl.gov/abs/0812.1151}{{\tt arXiv:0812.1151}}].

\bibitem{Eguchi:2009cq}
T.~Eguchi and K.~Hikami, {\it {Superconformal Algebras and Mock Theta Functions
  2. Rademacher Expansion for K3 Surface}},
  \href{http://xxx.lanl.gov/abs/0904.0911}{{\tt arXiv:0904.0911}}.

\bibitem{Eguchi:2010ej}
T.~Eguchi, H.~Ooguri, and Y.~Tachikawa, {\it {Notes on the K3 Surface and the
  Mathieu group M24}},  \href{http://xxx.lanl.gov/abs/1004.0956}{{\tt
  arXiv:1004.0956}}.

\bibitem{Cheng:2010pq}
M.~C.~N. Cheng, {\it {K3 Surfaces, N=4 Dyons, and the Mathieu Group M24}},
  \href{http://xxx.lanl.gov/abs/1005.5415}{{\tt arXiv:1005.5415}}.

\bibitem{Troost:2010ud}
J.~Troost, {\it {The non-compact elliptic genus: mock or modular}},  {\em JHEP}
  {\bf 06} (2010) 104, [\href{http://xxx.lanl.gov/abs/1004.3649}{{\tt
  arXiv:1004.3649}}].

\bibitem{Gottsche:1999}
L.~Gottsche, {\it {Theta Functions and Hodge Numbers of Moduli Spaces of
  Sheaves on Rational Surfaces}},  {\em Commun. Math. Phys.} {\bf 206} (1999)
  105--136.

\bibitem{Maldacena:1999bp}
J.~M. Maldacena, G.~W. Moore, and A.~Strominger, {\it {Counting BPS black holes
  in toroidal type II string theory}},
  \href{http://xxx.lanl.gov/abs/hep-th/9903163}{{\tt hep-th/9903163}}.

\bibitem{deBoer:2006vg}
J.~de~Boer, M.~C.~N. Cheng, R.~Dijkgraaf, J.~Manschot, and E.~Verlinde, {\it {A
  farey tail for attractor black holes}},  {\em JHEP} {\bf 11} (2006) 024,
  [\href{http://xxx.lanl.gov/abs/hep-th/0608059}{{\tt hep-th/0608059}}].

\bibitem{Kraus:2006nb}
P.~Kraus and F.~Larsen, {\it {Partition functions and elliptic genera from
  supergravity}},  {\em JHEP} {\bf 01} (2007) 002,
  [\href{http://xxx.lanl.gov/abs/hep-th/0607138}{{\tt hep-th/0607138}}].

\bibitem{Gaiotto:2007cd}
D.~Gaiotto and X.~Yin, {\it {Examples of M5-brane elliptic genera}},  {\em
  JHEP} {\bf 11} (2007) 004,
  [\href{http://xxx.lanl.gov/abs/hep-th/0702012}{{\tt hep-th/0702012}}].

\bibitem{Manschot:2007ha}
J.~Manschot and G.~W. Moore, {\it {A Modern Fareytail}},  {\em Commun. Num.
  Theor. Phys.} {\bf 4} (2010) 103--159,
  [\href{http://xxx.lanl.gov/abs/0712.0573}{{\tt arXiv:0712.0573}}].

\bibitem{Dabholkar:2005dt}
A.~Dabholkar, F.~Denef, G.~W. Moore, and B.~Pioline, {\it {Precision counting
  of small black holes}},  {\em JHEP} {\bf 10} (2005) 096,
  [\href{http://xxx.lanl.gov/abs/hep-th/0507014}{{\tt hep-th/0507014}}].

\bibitem{Minasian:1999qn}
R.~Minasian, G.~W. Moore, and D.~Tsimpis, {\it {Calabi-Yau black holes and
  (0,4) sigma models}},  {\em Commun. Math. Phys.} {\bf 209} (2000) 325--352,
  [\href{http://xxx.lanl.gov/abs/hep-th/9904217}{{\tt hep-th/9904217}}].

\bibitem{Guica:2007wd}
M.~Guica and A.~Strominger, {\it {Cargese lectures on string theory with eight
  supercharges}},  {\em Nucl. Phys. Proc. Suppl.} {\bf 171} (2007) 39--68,
  [\href{http://xxx.lanl.gov/abs/0704.3295}{{\tt arXiv:0704.3295}}].

\bibitem{Weist:2009}
T.~Weist, {\it Torus fixed points of moduli spaces of stable bundles of rank
  three},  \href{http://xxx.lanl.gov/abs/0903.0723}{{\tt arXiv:0903.0723}}.

\bibitem{Freed:1999vc}
D.~S. Freed and E.~Witten, {\it {Anomalies in string theory with D-branes}},
  \href{http://xxx.lanl.gov/abs/hep-th/9907189}{{\tt hep-th/9907189}}.

\bibitem{Minasian:1997mm}
R.~Minasian and G.~W. Moore, {\it {K-theory and Ramond-Ramond charge}},  {\em
  JHEP} {\bf 11} (1997) 002,
  [\href{http://xxx.lanl.gov/abs/hep-th/9710230}{{\tt hep-th/9710230}}].

\bibitem{Gottsche:1990}
L.~Gottsche, {\it {The Betti numbers of the Hilbert schemes of points on a
  smooth projective surface}},  {\em Math. Ann. 286} (1990) 193--207.

\bibitem{Joyce:2006pf}
Joyce, {\it {Holomorphic generating functions for invariants counting coherent
  sheaves on Calabi-Yau 3-folds}},
  \href{http://xxx.lanl.gov/abs/hep-th/0607039}{{\tt hep-th/0607039}}.

\bibitem{Joyce:2008}
S.~Joyce, {\it {A theory of generalized Donaldson-Thomas invariants}},
  \href{http://xxx.lanl.gov/abs/0810.5645}{{\tt 0810.5645}}.

\bibitem{Dabholkar:2005by}
A.~Dabholkar, F.~Denef, G.~W. Moore, and B.~Pioline, {\it {Exact and asymptotic
  degeneracies of small black holes}},  {\em JHEP} {\bf 0508} (2005) 021,
  [\href{http://xxx.lanl.gov/abs/hep-th/0502157}{{\tt hep-th/0502157}}].

\bibitem{Ganor:1996mu}
O.~J. Ganor and A.~Hanany, {\it {Small $E_8$ Instantons and Tensionless
  Non-critical Strings}},  {\em Nucl. Phys.} {\bf B474} (1996) 122--140,
  [\href{http://xxx.lanl.gov/abs/hep-th/9602120}{{\tt hep-th/9602120}}].

\bibitem{Seiberg:1996vs}
N.~Seiberg and E.~Witten, {\it {Comments on String Dynamics in Six
  Dimensions}},  {\em Nucl. Phys.} {\bf B471} (1996) 121--134,
  [\href{http://xxx.lanl.gov/abs/hep-th/9603003}{{\tt hep-th/9603003}}].

\bibitem{Morrison:1996na}
D.~R. Morrison and C.~Vafa, {\it {Compactifications of F-Theory on Calabi--Yau
  Threefolds -- I}},  {\em Nucl. Phys.} {\bf B473} (1996) 74--92,
  [\href{http://xxx.lanl.gov/abs/hep-th/9602114}{{\tt hep-th/9602114}}].

\bibitem{Morrison:1996pp}
D.~R. Morrison and C.~Vafa, {\it {Compactifications of F-Theory on Calabi--Yau
  Threefolds -- II}},  {\em Nucl. Phys.} {\bf B476} (1996) 437--469,
  [\href{http://xxx.lanl.gov/abs/hep-th/9603161}{{\tt hep-th/9603161}}].

\bibitem{Witten:1996qb}
E.~Witten, {\it {Phase Transitions In M-Theory And F-Theory}},  {\em Nucl.
  Phys.} {\bf B471} (1996) 195--216,
  [\href{http://xxx.lanl.gov/abs/hep-th/9603150}{{\tt hep-th/9603150}}].

\bibitem{Klemm:1996hh}
A.~Klemm, P.~Mayr, and C.~Vafa, {\it {BPS states of exceptional non-critical
  strings}},  \href{http://xxx.lanl.gov/abs/hep-th/9607139}{{\tt
  hep-th/9607139}}.

\bibitem{Lerche:1996ni}
W.~Lerche, P.~Mayr, and N.~P. Warner, {\it {Non-critical strings, del Pezzo
  singularities and Seiberg- Witten curves}},  {\em Nucl. Phys.} {\bf B499}
  (1997) 125--148, [\href{http://xxx.lanl.gov/abs/hep-th/9612085}{{\tt
  hep-th/9612085}}].

\bibitem{Minahan:1997ch}
J.~A. Minahan, D.~Nemeschansky, and N.~P. Warner, {\it {Investigating the BPS
  spectrum of non-critical E(n) strings}},  {\em Nucl. Phys.} {\bf B508} (1997)
  64--106, [\href{http://xxx.lanl.gov/abs/hep-th/9705237}{{\tt
  hep-th/9705237}}].

\bibitem{1-2-3-modular}
J.~Bruenier, G.~van~der Geer, G.~Harder, and D.~Zagier, {\em The 1-2-3 of
  Modular Forms}.
\newblock Springer, Universitext, 2008.

\bibitem{Yau:1995mv}
S.-T. Yau and E.~Zaslow, {\it {BPS States, String Duality, and Nodal Curves on
  K3}},  {\em Nucl. Phys.} {\bf B471} (1996) 503--512,
  [\href{http://xxx.lanl.gov/abs/hep-th/9512121}{{\tt hep-th/9512121}}].

\bibitem{MKPS}
A.~Klemm, D.~Maulik, R.~Pandharipande, and E.~Scheidegger, {\it
  Noether-lefschetz theory and the yau-zaslow conjecture},
  \href{http://xxx.lanl.gov/abs/0807.2477}{{\tt arXiv:0807.2477}}.

\bibitem{Hosono:1999qc}
S.~Hosono, M.~H. Saito, and A.~Takahashi, {\it {Holomorphic anomaly equation
  and BPS state counting of rational elliptic surface}},  {\em Adv. Theor.
  Math. Phys.} {\bf 3} (1999) 177--208,
  [\href{http://xxx.lanl.gov/abs/hep-th/9901151}{{\tt hep-th/9901151}}].

\bibitem{Hosono:2002xj}
S.~Hosono, {\it {Counting BPS states via holomorphic anomaly equations}},
  \href{http://xxx.lanl.gov/abs/hep-th/0206206}{{\tt hep-th/0206206}}.

\bibitem{Diaconescu:2007bf}
E.~Diaconescu and G.~W. Moore, {\it {Crossing the Wall: Branes vs. Bundles}},
  \href{http://xxx.lanl.gov/abs/0706.3193}{{\tt arXiv:0706.3193}}.

\bibitem{Maruyama:1977}
M.~Maruyama, {\it {Moduli of stable sheaves. II}},  {\em J. Math. Kyoto Univ.}
  {\bf 18} (1977) 557.

\bibitem{Yoshioka:1995}
K.~Yoshioka, {\it The betti numbers of the moduli space of stable sheaves of
  rank 2 on a ruled surface},  {\em Math. Ann.} (1995).

\bibitem{Yoshioka:1996}
K.~Yoshioka, {\it {The chamber structure of polarizations and the moduli of
  stable sheaves on a ruled surface}},  {\em Int. J. of Math.} {\bf 7} (1996)
  411--431, [\href{http://xxx.lanl.gov/abs/alg-geom/9409008}{{\tt
  alg-geom/9409008}}].

\bibitem{Li:1999}
W.-P. Li and Z.~Qin, {\it On blowup formulae for the s-duality conjecture of
  vafa and witten},  {\em Invent. Math.} {\bf 136} (1999) 451--482,
  [\href{http://xxx.lanl.gov/abs/math/9805054}{{\tt math/9805054}}].

\bibitem{Yoshioka:1994}
K.~Yoshioka, {\it The betti numbers of the moduli space of stable sheaves of
  rank 2 on $\mathbbm{P}^2$},  {\em J. reine angew. Math} {\bf 453} (1994).

\bibitem{Kool}
M.~Kool, {\it Euler characteristics of moduli spaces of torsion free sheaves on
  toric surfaces},  \href{http://xxx.lanl.gov/abs/0906.3393}{{\tt
  arXiv:0906.3393}}.

\bibitem{Yoshioka:1999}
K.~Yoshioka, {\it Euler characteristics of su(2) instanton moduli spaces on
  rational elliptic surfaces},  {\em Commun. Math. Phys.} {\bf 205} (1999)
  501--517.

\bibitem{Yoshioka:2008}
K.~Yoshioka, {\it Chamber structure of polarizations and the moduli of stable
  sheaves on a ruled surface},
  \href{http://xxx.lanl.gov/abs/alg-geom/9409008}{{\tt alg-geom/9409008}}.

\bibitem{Klyachko}
A.~Klyachko, {\it Moduli of vector bundels and numbers of classes,},  {\em
  Funct. Anal. and Appl.} {\bf 25} (1991) 67--68.

\bibitem{Zwegers:higherdepth}
S.~P. Zwegers, ``Mock modular forms.'' Talk given at the conference
  "Partitions, q-series and modular forms", University of Florida, Gainesville,
  March 12-16, 2008; Talk available under
  http://mathsci.ucd.ie/~zwegers/presentations/001.pdf.

\bibitem{Sen:2007vb}
A.~Sen, {\it {Walls of Marginal Stability and Dyon Spectrum in N=4
  Supersymmetric String Theories}},  {\em JHEP} {\bf 05} (2007) 039,
  [\href{http://xxx.lanl.gov/abs/hep-th/0702141}{{\tt hep-th/0702141}}].

\bibitem{Cheng:2007ch}
M.~C.~N. Cheng and E.~Verlinde, {\it {Dying Dyons Don't Count}},  {\em JHEP}
  {\bf 09} (2007) 070, [\href{http://xxx.lanl.gov/abs/0706.2363}{{\tt
  arXiv:0706.2363}}].

\bibitem{Nekrasov:2002qd}
N.~A. Nekrasov, {\it {Seiberg-Witten Prepotential From Instanton Counting}},
  {\em Adv. Theor. Math. Phys.} {\bf 7} (2004) 831--864,
  [\href{http://xxx.lanl.gov/abs/hep-th/0206161}{{\tt hep-th/0206161}}].

\bibitem{Iqbal:2007ii}
A.~Iqbal, C.~Kozcaz, and C.~Vafa, {\it {The refined topological vertex}},  {\em
  JHEP} {\bf 10} (2009) 069,
  [\href{http://xxx.lanl.gov/abs/hep-th/0701156}{{\tt hep-th/0701156}}].

\bibitem{Goettsche:2007}
L.~Gottsche, H.~Nakajima, and K.~Yoshioka, {\it K-theoretic donaldson
  invariants via instanton counting},
  \href{http://xxx.lanl.gov/abs/math/0611945}{{\tt math/0611945}}.

\bibitem{Ooguri:2004zv}
H.~Ooguri, A.~Strominger, and C.~Vafa, {\it {Black hole attractors and the
  topological string}},  {\em Phys. Rev.} {\bf D70} (2004) 106007,
  [\href{http://xxx.lanl.gov/abs/hep-th/0405146}{{\tt hep-th/0405146}}].

\bibitem{Aganagic:2009kf}
M.~Aganagic, H.~Ooguri, C.~Vafa, and M.~Yamazaki, {\it {Wall Crossing and
  M-theory}},  \href{http://xxx.lanl.gov/abs/0908.1194}{{\tt arXiv:0908.1194}}.

\bibitem{David:2006ru}
J.~R. David, D.~P. Jatkar, and A.~Sen, {\it {Dyon spectrum in N = 4
  supersymmetric type II string theories}},  {\em JHEP} {\bf 11} (2006) 073,
  [\href{http://xxx.lanl.gov/abs/hep-th/0607155}{{\tt hep-th/0607155}}].

\bibitem{delPezzo}
V.~Alexeev and V.~V. Nikulin, {\it Del pezzo and k3 surfaces},  {\em
  Mathematical Society of Japan Memoirs} {\bf 15} (2006) 1--164.

\bibitem{Aspinwall:2004jr}
P.~S. Aspinwall, {\it {D-branes on Calabi-Yau manifolds}},
  \href{http://xxx.lanl.gov/abs/hep-th/0403166}{{\tt hep-th/0403166}}.

\bibitem{Green:1996dd}
M.~B. Green, J.~A. Harvey, and G.~W. Moore, {\it {I-brane inflow and anomalous
  couplings on D-branes}},  {\em Class. Quant. Grav.} {\bf 14} (1997) 47--52,
  [\href{http://xxx.lanl.gov/abs/hep-th/9605033}{{\tt hep-th/9605033}}].

\bibitem{Witten:1998cd}
E.~Witten, {\it {D-branes and K-theory}},  {\em JHEP} {\bf 12} (1998) 019,
  [\href{http://xxx.lanl.gov/abs/hep-th/9810188}{{\tt hep-th/9810188}}].

\bibitem{Diaconescu:1999vp}
D.-E. Diaconescu and C.~Romelsberger, {\it {D-branes and bundles on elliptic
  fibrations}},  {\em Nucl. Phys.} {\bf B574} (2000) 245--262,
  [\href{http://xxx.lanl.gov/abs/hep-th/9910172}{{\tt hep-th/9910172}}].

\bibitem{Douglas:2000gi}
M.~R. Douglas, {\it {D-branes, categories and N = 1 supersymmetry}},  {\em J.
  Math. Phys.} {\bf 42} (2001) 2818--2843,
  [\href{http://xxx.lanl.gov/abs/hep-th/0011017}{{\tt hep-th/0011017}}].

\bibitem{Douglas:2000ah}
M.~R. Douglas, B.~Fiol, and C.~Romelsberger, {\it {Stability and BPS branes}},
  {\em JHEP} {\bf 09} (2005) 006,
  [\href{http://xxx.lanl.gov/abs/hep-th/0002037}{{\tt hep-th/0002037}}].

\bibitem{Manschot:2008zb}
J.~Manschot, {\it {On the space of elliptic genera}},  {\em Commun. Num. Theor.
  Phys.} {\bf 2} (2008) 803--833,
  [\href{http://xxx.lanl.gov/abs/0805.4333}{{\tt arXiv:0805.4333}}].

\bibitem{Hecht:2008}
M.~Hecht, {\it {Black Holes in M-Theory, BPS states and modularity}},  {\em
  Diploma thesis at the Ludwig-Maximilians University of Munich} (2008).

\end{thebibliography}\endgroup
\bibliographystyle{JHEP}

\end{document}